\documentclass[aps,pra,preprint,superscriptaddress]{revtex4-1}
\usepackage{graphicx}
\usepackage{amsmath}
\usepackage{amssymb}

\newcommand{\Eq}[1]{Eq.(\ref{#1})}% \Eq{eq:abc}
\newcommand{\Fig}[1]{Fig.\,\ref{#1}}% \Fig{fig:abc}
\newcommand{\be}{\begin{equation}}
\newcommand{\ee}{\end{equation}}
\newcommand{\bea}{\begin{eqnarray}}
\newcommand{\eea}{\end{eqnarray}}

\newcommand{\bp}{\mathbf{p}}
\newcommand{\bE}{\mathbf{E}}
\newcommand{\br}{\mathbf{r}}
\newcommand{\epsm}{\epsilon_{\rm m}}

\newcommand{\alphah}{\hat{\alpha}}

\newcommand{\EF}{E_{\rm FWM}}
\newcommand{\AF}{A_{\rm FWM}}
\newcommand{\PhiF}{\Phi_{\rm FWM}}

\newcommand{\Er}{E_{\rm 2r}}
\newcommand{\Ar}{A_{\rm 2r}}
\newcommand{\Phir}{\Phi_{\rm 2r}}

\newcommand{\am}{a_{\rm m}}
\newcommand{\omm}{\omega_{\rm m}}

\newcommand{\Pref}{P_{\rm R}}
\newcommand{\Pel}{P_{\rm el}}

\newcommand{\RNP}{R_{\rm P}}
\newcommand{\RC}{R_{\rm C}}
\newcommand{\tauC}{\tau_{\rm C}}

\newcommand{\kB}{k_{\rm B}}

\begin{document}

%\preprint{}

\title{Background-free 3D nanometric localisation and sub-nm asymmetry detection of single plasmonic nanoparticles by four-wave mixing interferometry with optical vortices}



\author{George Zoriniants}
\affiliation{Cardiff University School of Biosciences, Museum
Avenue, Cardiff CF10 3AX, United Kingdom}
\author{Francesco Masia}
\affiliation{Cardiff University School of Physics and Astronomy, The
Parade, Cardiff CF24 3AA, United Kingdom}
\author{Naya Giannakopoulou}
\affiliation{Cardiff University School of Biosciences, Museum
Avenue, Cardiff CF10 3AX, United Kingdom}
\author{Wolfgang Langbein}
\email{langbeinww@cardiff.ac.uk} \affiliation{Cardiff University
School of Physics and Astronomy, The Parade, Cardiff CF24 3AA,
United Kingdom}
\author{Paola Borri}
\email{borrip@cardiff.ac.uk}
\affiliation{Cardiff University School of Biosciences, Museum Avenue, Cardiff CF10 3AX, United Kingdom}


\date{\today}

\begin{abstract}
Single nanoparticle tracking using optical microscopy is a powerful
technique with many applications in biology, chemistry and material
sciences. Despite significant advances, localising objects with
nanometric position accuracy in a scattering environment remains
challenging. Applied methods to achieve contrast are dominantly
fluorescence based, with fundamental limits in the emitted photon
fluxes arising from the excited-state lifetime as well as
photobleaching. Furthermore, every localisation method reported to
date requires signal acquisition from multiple spatial points, with
consequent speed limitations. Here, we show a new four-wave mixing
interferometry technique, whereby the position of a single
non-fluorescing gold nanoparticle is determined with better than
20\,nm accuracy in plane and 1\,nm axially from rapid single-point
acquisition measurements by exploiting optical vortices. The
technique is also uniquely sensitive to particle asymmetries of only
0.5\% aspect ratio, corresponding to a single atomic layer of gold,
as well as particle orientation, and the detection is
background-free even inside biological cells. This method opens new
ways of of unraveling single-particle trafficking within complex 3D
architectures.
\end{abstract}

%\pacs{}

\maketitle

The ability to localize optically and eventually track the position
of objects at the nanoscale requires ways to overcome the Abbe
diffraction limit given by $\lambda/(2{\rm NA}$), where $\lambda$ is
the wavelength of light and NA is the numerical aperture of the
imaging lens. For visible light and high NA objectives this limit is
roughly 200\,nm.

Current techniques for overcoming this limit can be divided into
near-field and far-field optical methods. In near-field optics,
super-resolution is achieved by localizing the light field (incident
and/or detected) using optical fiber probes with sub-wavelength
apertures\,\cite{BetzigScience92}, metal coated
tips\,\cite{SanchezPRL99}, or plasmonic
nano-antennas\,\cite{SchuckPRL05}. These techniques can provide
spatial resolutions down to about 10\,nm, however are limited to
interrogating structures accessible to optical tips and/or patterned
substrates and require precise position control of the probe.
Far-field methods overcome this drawback and allow to localise
single biomolecules inside cells. However, to achieve sufficient
contrast and specificity against backgrounds, super-resolution
far-field field methods are dominantly fluorescence-based. They
exploit either the principle of single emitter
localisation\,\cite{RustNatMeth06,BetzigScience06} or optical
point-spread function (PSF) engineering\,\cite{HellScience07} and
the fluorophore non-linear response in absorption or stimulated
emission. Beside specific advantages and disadvantages of these two
approaches, their implementation using fluorescence results in
significant limitations. Fluorophores are single quantum emitters
and thus only capable of emitting a certain maximum number of
photons per unit time due to the finite duration of their
excited-state lifetime, moreover they are prone to photobleaching
and associated photo-toxicity.

Alternatively to using fluorescent emitters, far-field nanoscopy
techniques have been shown with single metallic nanoparticles (NPs)
optically detected owing to their strong scattering and absorbtion
at the localised surface plasmon resonance (LSPR). These detection
methods are photostable, and the achievable photon fluxes are
governed by the incident photon fluxes and the NP optical extinction
cross-section. Position localisation below diffraction similar to
single emitter localisation can be achieved via wide-field
techniques such as bright-field microscopy\,\cite{FujiwaraJCB02},
dark-field microscopy\,\cite{UenoBJ10}, differential interference
contrast\,\cite{GuiAC12}, and interferometric scattering
microscopy\,\cite{Ortega-ArroyoPCCP12}. These techniques, however,
are not background--free and either use large NPs (diameters
$\geq40$\,nm) to distinguish them against endogenous scattering and
phase contrast in heterogeneous samples or work in optically clear
environments. A more selective technique uses photothermal imaging
where the image contrast originates from the refractive index change
in the region surrounding the nanoparticle due to local heating
following light absorption. This is a focused-beam scanning
technique and has been used to track single 5\,nm diameter gold NPs
in two dimensions (however not 3D)\,\cite{LasneBJ06}. By detecting
the nanoparticle only indirectly via the photothermal index change
generated in its surrounding, this method is also not free from
backgrounds. In fact, photothermal contrast has been shown in the
absence of NPs due to endogeneous absorption in
cells\,\cite{LasneOE07}. Moreover, as for every localisation method,
it required signal acquisition from multiple spatial points to
determine the position of the single NP, with consequent limitation
in tracking speed.

Four-wave mixing (FWM), triply-resonant to the LSPR, was shown by us
to be a very selective, high-contrast photostable method to detect
single small gold NPs\,\cite{MasiaOL09,MasiaPRB12}. It is a
third-order nonlinearity which originates from the change in the NP
dielectric constant induced by the resonant absorption of a pump
pulse and subsequent formation of a non-equilibrium hot electron gas
in the metal\,\cite{MasiaPRB12}. It is therefore very specific to
metallic NPs which are imaged background-free even in highly
scattering and fluorescing environments\,\cite{MasiaOL09}. In this
work, we show theoretically and experimentally a new FWM detection
modality which enables to determine the position of a single gold NP
with $<20$\,nm accuracy in plane and $<1$\,nm axially from scanless
single-point background-free acquisition in the 1\,ms time scale, by
exploiting optical vortices of tightly focussed light.

%\begin{section}{Results}
\section{Four-wave mixing interferometry technique}

\begin{figure*}
    \includegraphics*[width=13cm]{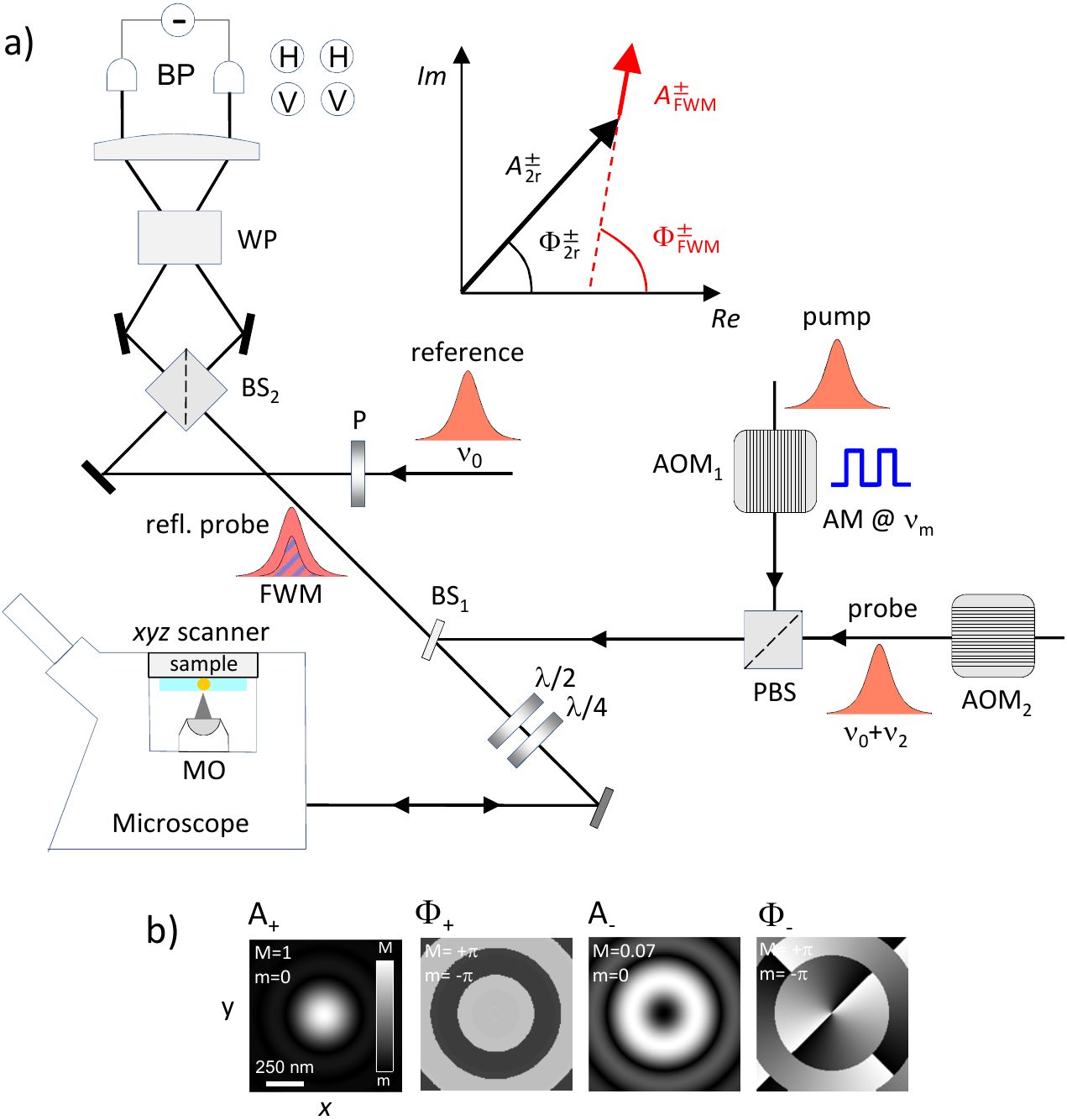}
    \caption{{\bf Four-wave mixing interferometry epi-detected dual-polarisation resolved.} a) Sketch. Pump pulses are amplitude-modulated
        (AM) at $\nu_{\rm m}$ and probe pulses are frequency shifted by
        $\nu_{2}$, using acousto-optical modulators (AOMs). Pulses are
        coupled into an inverted microscope equipped with a high NA
        microscope objective (MO). Pump and probe beams are adjusted to be
        circularly polarized at the sample by $\lambda/4$
        and $\lambda/2$ waveplates. Circular polarisations are transformed
        into horizontal (H) and vertical (V) linear polarisation by the same waveplates and both components are simultaneously detected through their interference with a frequency-unshifted reference linearly polarised at $45^\circ$. BP: Balanced photodiodes. WP: Wollaston prism deflecting beam out of drawing plane. (P)BS: (polarising) beam splitter. P: Polariser. Inset:
        amplitude ($A$) and phase ($\Phi$) of the reflected probe and FWM field, measured by a multi-channel lock-in, where +(-) refers to the co(cross) polarised
        component relative to the incident circularly polarized probe. b)
        Calculated field distribution in the focal plane of a 1.45\,NA
        objective for an incident field left circularly polarized. $A_+$ $(A_-)$ is the co (cross) polarised amplitude , and $\Phi_+$ ($\Phi_-$) the corresponding phase.\label{figSetUp}}
\end{figure*}

A sketch of the FWM technique is shown in Fig.\,\ref{figSetUp}a. The
key developments in this new design compared to previous works are
the epi-collection (reflection) geometry and the dual-polarisation
heterodyne detection scheme. We use a train of femtosecond optical
pulses with repetition rate $\nu_{\rm L}$ which is split into three
beams, all having the same center wavelength, resulting in a triply
degenerate FWM scheme. One beam acts as pump and excites the NP at
the LSPR with an intensity which is temporally modulated with close
to unity contrast by an acousto-optic modulator (AOM$_1$) driven at
carrier frequency $\nu_1$ with a square wave amplitude modulation of
frequency $\nu_{\rm m}$. The change in the NP optical properties
induced by this excitation is resonantly probed by a second pulse at
an adjustable delay time $\tau$ after the pump pulse. Pump and probe
pulses of fields ${\bE}_1$ and ${\bE}_2$, respectively, are
recombined into the same spatial mode and focused onto the sample by
a high NA microscope objective. The sample can be positioned and
moved with respect to the focal volume of the objective by scanning
an $xyz$ sample stage with nanometric position accuracy. A FWM field
${\bE}_{\rm FWM}$ (proportional to
${\bE}_{1}{\bE}_{1}^{*}{\bE}_{2}$) is collected together with the
probe in reflection (epi-direction) by the same objective,
transmitted by the beam splitter (BS$_1$) used to couple the
incident beams onto the microscope, and recombined in a second beam
splitter (BS$_2$) with a reference pulse field (${\bE}_{\rm R}$) of
adjustable delay. The resulting interference is detected by two
pairs of balanced photodiodes. A heterodyne scheme discriminates the
FWM field from pump and probe pulses and detects amplitude and phase
of the field. In this scheme, the probe optical frequency is
slightly upshifted via a second AOM (AOM$_2$), driven with a
constant amplitude at a radiofrequency of $\nu_{2}$, and the
interference of the FWM with the unshifted reference field is
detected. As a result of the amplitude modulation of the pump at
$\nu_{\rm m}$ and the frequency shift of the probe at $\nu_2$, this
interference gives rise to a beat note at $\nu_2$ with two sidebands
at $\nu_2\pm\nu_{\rm m}$, and replica separated by the repetition
rate $\nu_{\rm L}$ of the pulse train. A multi-channel lock-in
amplifier enables the simultaneous detection of the carrier at
$\nu_2-\nu_{\rm L}$ and the sidebands at $\nu_2\pm\nu_{\rm
m}-\nu_{\rm L}$. Via the in-phase ($Re$) and in-quadrature ($Im$)
components for each detected frequency, amplitude and phase of the
probe field reflected by the sample $({\bE}_{\rm 2r})$ and of the
epi-detected FWM field are measured (see sketch in
Fig.\,\ref{figSetUp}a).

A key point of the technique is the use of a dual-polarisation
balanced detection. Firstly, probe and pump beams, linearly
polarised horizontally (H)  and vertically (V) respectively in the
laboratory system, are transformed into cross-circularly polarized
beams at the sample by a combination of $\lambda/4$ and $\lambda/2$
waveplates. The reflected probe and FWM fields collected by the same
microscope objective travel backwards through the same waveplates,
such that the probe reflected by a planar surface returns V polarized in the laboratory system.
The reference beam is polarised at 45 degree (using a polariser) prior recombining with the
epi-detected signal via the non-polarizing beamsplitter BS$_2$. A
Wollaston prism vertically separates H and V polarizations for each
arm of the interferometer after BS$_2$. Two pairs of balanced
photodiodes then provide polarization resolved detection, the bottom
(top) pair detecting the current difference (for common-mode noise
rejection) of the V (H) polarised interferometer arms. In turn, this
corresponds to detecting the co- and cross-circularly polarised
components of ${\bE}_{\rm 2r}$ and ${\bE}_{\rm FWM}$ relative to the
incident circularly polarized probe, having amplitudes (phases)
indicated as $\Ar^\pm$ and $\AF^\pm$ ($\Phir^\pm$ and $\PhiF^\pm$)
in the sketch in \Fig{figSetUp}a where $+$ ($-$) refers to the co
(cross) polarised component.

\section{The concept of nanometric localisation using optical
vortices}

To elucidate conceptually how the nanometric position accuracy
arises from this dual-polarisation resolved FWM interferometry
detection scheme, we simulated numerically the field
distribution in the focal region of a 1.45\,NA objective. The
simulation parameters (wavelength, coverslip thickness, medium
refractive index, back objective filling factor) were chosen to
match the actual experimental conditions (see Methods). The
amplitude and phase components of the field in the focal plane are
shown in \Fig{figSetUp}b for a left circularly ($\sigma^+$)
polarised  input field. Due to the high NA of the objective and the
vectorial nature of the field, there is a significant
cross-circularly polarised component that forms an optical
vortex of topological charge $l=2$, i.e. it  has an amplitude
($A_-$) which is zero in the focus center and radially-symmetric
non-zero away from the center, and a phase ($\Phi_-$) changing with
twice the in-plane polar angle. A point-like gold NP displaced from
the focus center experiences this field distribution,
and will in turn emit a field with an amplitude and a phase directly related to the NP radial and angular position.
While this is the case for both reflected and FWM fields, only the FWM signal is background-free hence suited for tracking the NP in heterogeneous environments. In the following,
we discuss the FWM field emitted from such NP and
the corresponding coordinate retrieval.

We calculate the FWM field starting from the polarization of the NP
at position $\br$ induced by the probe field, described by
$\bp(\br)=\epsilon_0 \epsm \alphah \bE_2(\br)$ where $\epsilon_0$ is
the vacuum permittivity, $\epsm$ is the dielectric constant of the
medium surrounding the NP (glass and silicon oil in the experiment)
and $\alphah$ is the particle polarizability tensor. We calculate
$\alphah$ in the dipole approximation, valid for particle sizes much
smaller than the wavelength of light (Rayleigh regime). To take into
account non-sphericity of real particles, which will be
relevant in the experiments as shown later, we adopted the model of
a metallic ellipsoid with three orthogonal semi-axes of symmetry
$a$, $b$ and $c$. In the particle reference system, $\alphah$ is
diagonal and its eigenvalues are given
by\,\cite{bohren1998absorption} $\alpha_i= 4 \pi abc
(\epsilon-\epsm)/(3\epsm+3L_i(\epsilon-\epsm))$, where $\epsilon$ is
the dielectric permittivity of the particle, and $L_i$ with
$i=a,b,c$  are dimensionless quantities defined by the particle
geometry (see Supplementary Information section S1.i). For an
arbitrary particle orientation in the laboratory system, the
polarizability tensor can be transformed using
$\alphah=\hat{M}^{-1}\alphah'\hat{M}$ where $\hat{M}$ is a rotation
matrix.

\begin{figure*}
\includegraphics*[width=16.5cm]{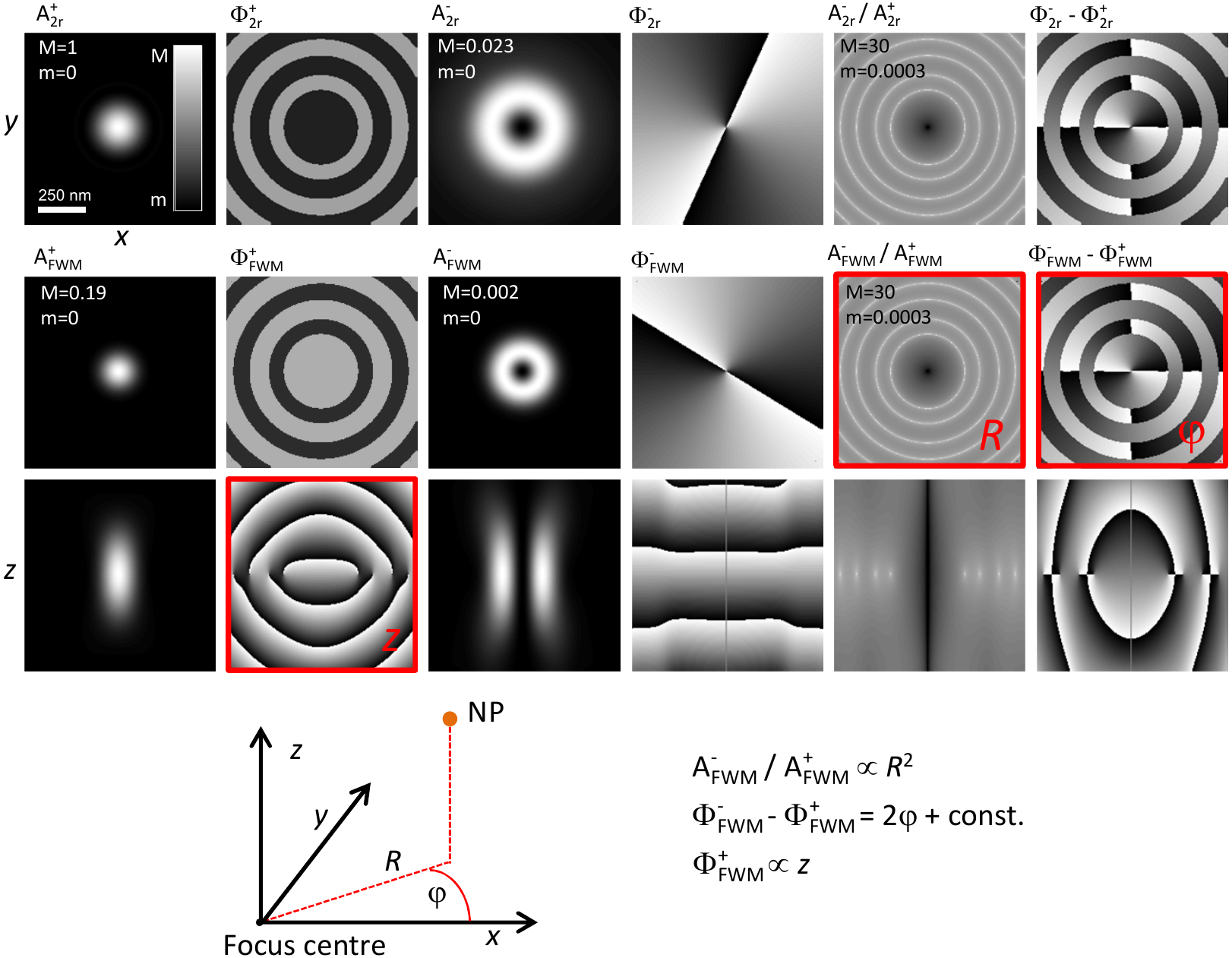}
\caption{{\bf Nanometric localisation using optical vortices.} Calculated amplitude (phase) components of the reflected probe
field and FWM field $\Ar^\pm$ and $\AF^\pm$ ($\Phir^\pm$ and $
\PhiF^\pm$) respectively, as a function of particle position in the sample focal plane $(x,y)$,
and in a section along the axial direction $(x,z)$ through the
focus, where + refers to the co-polarised
component and - to the cross-polarised component relative to the
left-circularly polarised incident probe. The calculation assumes a
perfectly spherical gold NP in the dipole approximation. The inset
shows a sketch of how the amplitude and phase of the FWM field ratio
and the phase of the co-polarised FWM field can be used to locate in
3D the spatial position of the NP relative to the focus center.
Linear grey scale from $-\pi$ to $\pi$ for all phases, and from m to
M for field amplitudes, as indicated. The amplitude ratio of
reflected probe and FWM are shown on a logarithmic greyscale over 5
orders of magnitudes. \label{figSim}}
\end{figure*}

The interference of the reflected probe field with the reference
field is calculated using $ E_{\rm 2r}^\pm=(\bE_{\rm R}^\pm)^*\cdot
\alphah\,{\bE_2^+}$ where $\alphah\,{\bE_2^+}$ is the particle
induced polarisation for a $\sigma^+$ polarised incident probe field
(we dropped the constant $\epsilon_0\epsm$ in $\bp$ for brevity),
and $\bE_{\rm R}^\pm$ are reference fields equal to left (+) and right
(-) circularly polarised input fields. The technique is configured
such that the optical mode of probe and reference fields are
matched, hence $\bE_{\rm R}$ was calculated as the field
distribution in the focal region in the same way as $\textbf{E}_{2}$
(see Methods), and back-propagated via time reversal. Similarly, the FWM interference is calculated as
$E_{\rm FWM}^\pm =(\bE_{\rm R}^\pm)^*\cdot\delta\alphah\,{\bE_2^+}$.
Here, $\delta\alphah$ is the pump-induced change of the particle
polarizability which we have modeled as described in our previous
work\,\cite{MasiaPRB12}. Briefly, $\delta\alphah$ arises from the
transient change of the electron and lattice temperature following
the absorption of the pump pulse by the NP. $\delta\alphah$ depends
on the pump fluence at the NP, on the particle absorption
cross-section, and on the delay time between pump and probe pulses
(see also Supplementary Information Fig.\,S2 and S10). The
simulations in Fig.\,\ref{figSim} were performed to reproduce the
experimentally measured FWM signal strength on a 30\,nm radius gold
NP at $\tau=0.5$\,ps shown later (note that $\tau\sim0.5$\,ps is the
delay for which the FWM amplitude reaches its maximum as a result of
the ultrafast heating of the electron gas\,\cite{MasiaPRB12}).

$\Er^\pm$ and $\EF^\pm$ have amplitudes and phases as a function of particle position in the focal region as shown in
Fig.\,\ref{figSim} for the case of a perfectly spherical NP. Similar
to the spatial distributions of the focussed field shown in Fig.\,\ref{figSetUp}b, $\Er^-$ and $\EF^-$
form optical vortices of $l=2$ topological charge. The ratio $\Er^- /
\Er^+$ is also shown with its amplitude
$\Ar^- / {\Ar^+}$ and phase $\Phir^- - \Phir^+$. The FWM field
distribution has a narrower PSF than the reflection, as expected from the third-order
nonlinearity. The phase of $\EF$ is shifted compared to $\Er$ due to
the phase difference between $\delta\alphah$ and $\alphah$. Slices
along the axial direction ($z$) are shown for $\EF$ (a similar
behavior is observed for $\Er$, see Supplementary Information
Fig.\,S1). Of specific importance for the localization of the NP in
3D are the amplitude and phase of the FWM field ratio $\EF^- /
\EF^+$ and the phase of the co-polarized $\EF^+$ as highlighted by
the red panels and sketch in \Fig{figSim}.

The co-polarized FWM amplitude is much stronger ($\sim90$ fold) than
the cross-polarized FWM, and has a phase which is independent of the
lateral position over the PSF width. Conversely, along the $z$-axis this phase
is linear in the displacement between particle and center of the
focus and can be used to determine the particle $z$  coordinate.
This can be easily understood as due to the optical path length
difference between the particle and the observation point. For a
plane wave of wavevector $k=2{\pi}n/\lambda$ with $n$ refractive
index in the medium the phase would be $2kz$, the factor of 2 accounting
for the back and forth path in reflection geometry reflection. We find a linear relationship between $\PhiF^+$ and
$z$ (see Supplementary Information Fig.\,S3)  with a slope $\partial
z/\partial\Phi=38.8$\,nm/rad, slightly larger than $\lambda/(4\pi
n)=28.8$\,nm/rad. This is due to the propagation of a focussed beam
with high NA where a Gouy phase shift occurs, reducing the wavevector
in axial direction due to the wavevector spread in lateral direction.

For the in-plane radial coordinate $R$ of the NP position relative
to the focus position (see sketch in \Fig{figSim}), we find that the
FWM amplitude ratio $\AF^-/\AF^+$ scales quadratically with $R$ up
to $R\sim60$\,nm, such that this coordinate can be calculated as
$R=R_{0}\sqrt{\AF^-/\AF^+}$. Notably, by using the FWM {\it ratio}
the retrieved $R$ is independent of the pump, probe
and reference powers, and of the NP size (in the dipole
approximation). Conversely, $R_{0}$ is specific to the NA of the
microscope objective used and the probe beam fill factor (see Supplementary Information Fig.\,S4)
and decreases with increasing NA, showing that high NAs are required
to localize the NP in $R$ down to small distances. Finally, the
angular position coordinate $\varphi$ can be taken from the phase of
the FWM ratio $\Theta=\PhiF^- - \PhiF^+$ as
$\varphi=(\Theta-\Theta_0)/2$ (see also Supplementary Information
Fig.\,S3).

In essence, using these relationships we can locate the NP in 3D
($z$, $R$ and $\varphi$) by scanless polarisation-resolved and
phase-resolved FWM acquisition at a single spatial point.

\section{Background-free Four-wave Mixing Detection}

\begin{figure}
\includegraphics*[width=16cm]{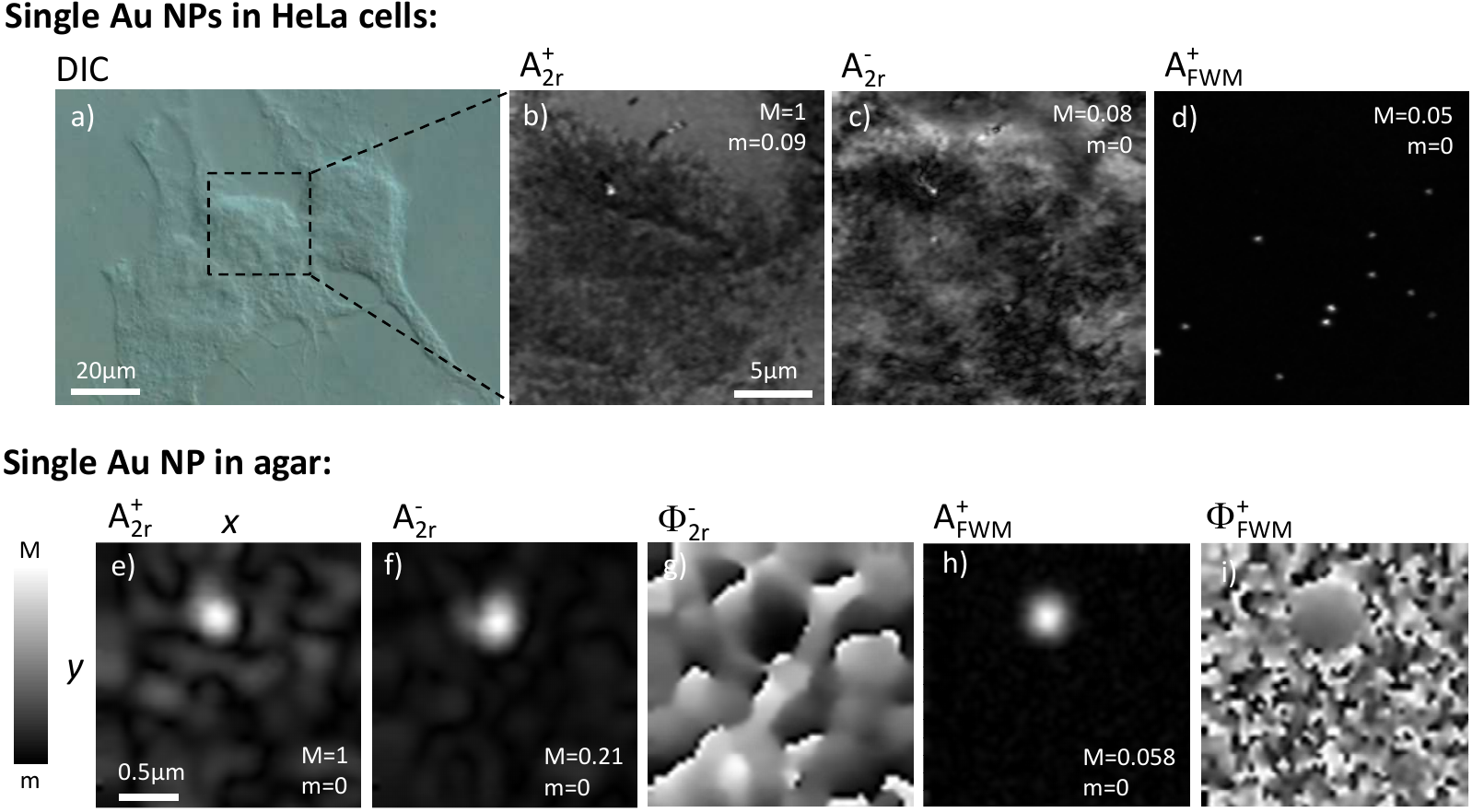}
\caption{{\bf Background-free FWM detection of single NPs in
heterogeneous environments.} Top: Fixed HeLa cells that have
internalized gold NPs of 20\,nm radius imaged by (a) differential
interference contrast (DIC) microscopy, (b) co-circularly polarised
reflection amplitude $\Ar^+$, (c) cross-circularly polarised
reflection amplitude $\Ar^-$, (d) co-circularly polarised FWM
amplitude $\AF^+$. FWM was acquired with pump-probe delay time of
0.5\,ps, pump (probe) power at the sample of 30\,$\mu$W
(15\,$\mu$W), 2\,ms pixel dwell time, pixel size in plane of 95\,nm
and z-stacks over 6\,$\mu$m in 250\,nm z-steps. FWM is shown as a
maximum intensity projection over the z-stack, while the reflection
is on a single $(x,y)$ plane (scanning the sample position). Bottom:
Single $(x,y)$ plane image of a 25\,nm radius gold NP in 5\% agarose
gel, via the co-circularly polarised reflection amplitude $\Ar^+$
(e), cross-circularly polarised reflection amplitude $\Ar^-$ (f) and
phase $\Phir^-$ (g), co-circularly polarised FWM amplitude $\AF^+$
(h) and phase $\PhiF^+$ (i). FWM was acquired with 0.5\,ms pixel
dwell time, pixel size in plane of 38\,nm. Grey scales are linear
from $-\pi$ to $\pi$ for all phases, and from m to M for field
amplitudes, as indicated.} \label{figcell}
\end{figure}

Prior to quantifying experimentally the nanometric localisation
accuracy, it is important to emphasize that our FWM detection is
background-free even in scattering and/or autofluorescing
environments, making it applicable to imaging single small NPs inside cells, surpassing other methods
reported in the literature. Localizing single plasmonic NPs with
nanometric accuracy even at sub-millisecond exposure times can be
achieved in an optically clear environment with simpler techniques
such as dark-field microscopy\,\cite{UenoBJ10} and interferometric
scattering microscopy\,\cite{Ortega-ArroyoPCCP12}. However, since these
techniques use the linear response of the NP they are substantially affected by endogenous scattering and
fluorescence, which severely limit their practical applicability in
heterogeneous biological environments.

To exemplify this point, Fig.\,\ref{figcell}(top) shows fixed HeLa
cells that have internalized gold NPs of 20\,nm radius, imaged with
FWM using a 1.45\,NA oil-immersion objective. High-resolution DIC
microscopy was available in the same instrument (for details see
Supplementary Information section S2.iii and S2.vii).
Fig.\,\ref{figcell}a shows the DIC image of a group of HeLa cells on
which reflection and FWM imaging was performed in the region
highlighted by the dashed frame. The co-circularly polarized
reflection image $\Ar^+$ shown in Fig.\,\ref{figcell}b correlates
with the cell contour seen in DIC, and shows a spatially varying
contrast due to thickness and refractive index inhomogeneities in
the sample. Even with a particle diameter as large as 40\,nm, gold
NPs are not distinguished from the cellular contrast neither in DIC
nor in the $\Ar^+$ reflection image. Detecting the cross-polarized
reflection $\Ar^-$ (Fig.\,\ref{figcell}c) which has been suggested
as a way to improve contrast\,\cite{MilesACSPhotonics15} is still
severely affected by the cellular scattering background. On the
contrary, the co-circularly polarised FWM amplitude $\AF^+$ shown in
Fig.\,\ref{figcell}d as maximum intensity projection over a
6\,${\mu}$m $z$-stack is background-free (throughout the $z$-stack)
and clearly indicates the location of single gold NPs in the cell.
Notably, FWM acquisition can be performed simultaneously with
confocal fluorescence microscopy for correlative co-localisation
analysis, as shown in the Supplementary Information Fig.\,S11.

Fig.\,\ref{figcell}(bottom) shows a single 25\,nm radius gold NP
embedded in a dense (5\% w/v) agarose gel. The scattering from the
gel is visible as a structured background in the co-circularly
polarized reflection image $\Ar^+$ in Fig.\,\ref{figcell}e.
Detecting the cross-polarized reflection (Fig.\,\ref{figcell}f)
again does not eliminate the background. With the reflection
amplitude scaling as the volume of the particle, a 15\,nm radius NP
would be indistinguishable from the background in the present case.
Furthermore, the interference with the background limits the
accuracy of particle localization. Notably, the phase $\Phir^-$ is
also severely affected by the scattering from the gel. Conversely,
the FWM amplitude $\AF^+$ and phase $\PhiF^+$ shown in
Fig.\,\ref{figcell}h,i are background-free (as can be seen by the
random phase outside the particle) and clearly resolve the NP
despite the heterogeneous surrounding.

\section{Experimental localisation: The role of nanoparticle
ellipticity}

To experimentally quantify the nanometric localisation and its
accuracy due to photon shot-noise, we started by examining a single
nominally spherical 30\,nm radius gold NP drop cast onto a glass
coverslip and immersed in silicon oil, using an index-matched
1.45\,NA oil-immersion objective.

\begin{figure*}
\includegraphics*[height=18cm]{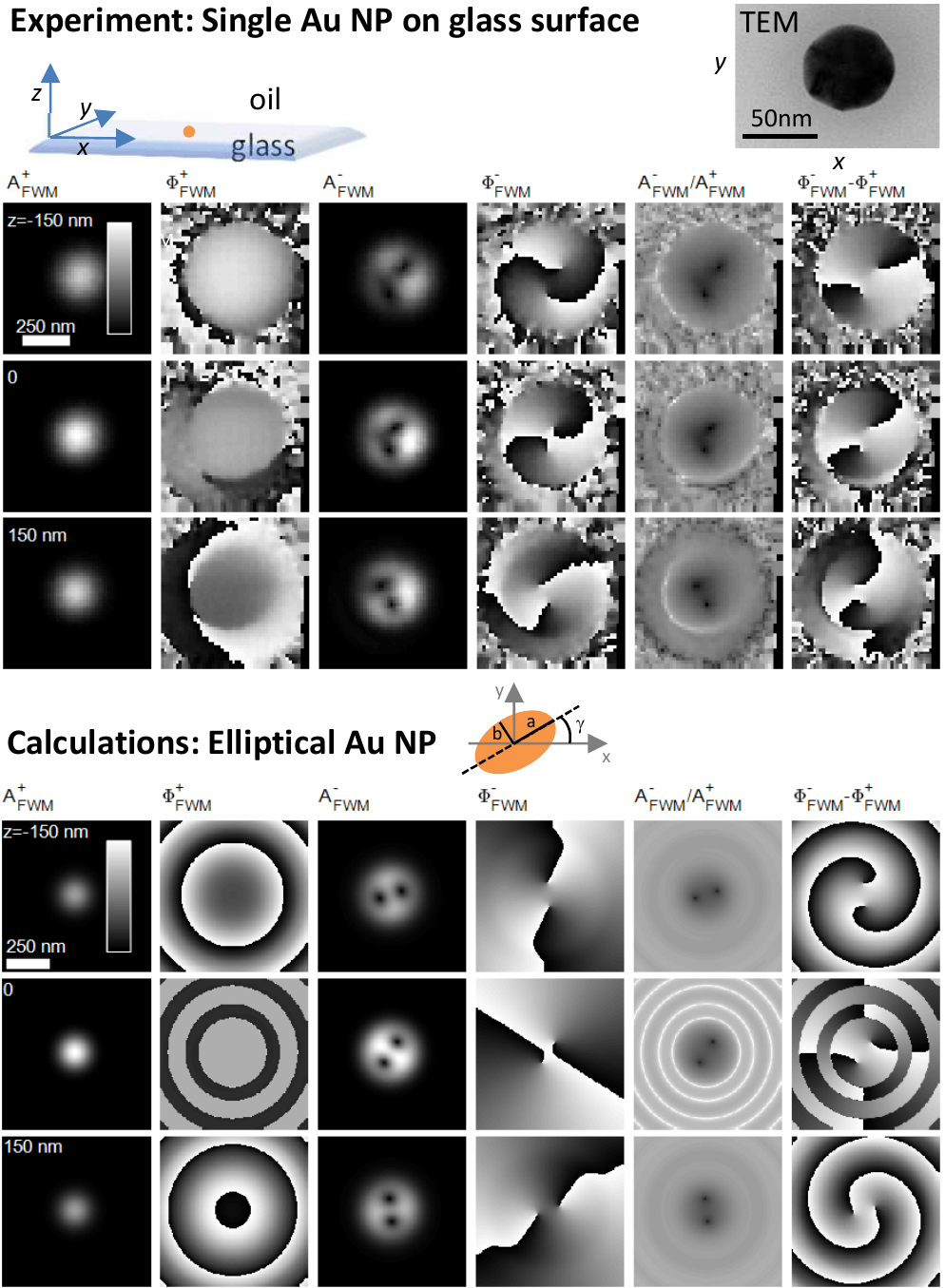}
\caption{{\bf FWM of a single NP attached on glass.} Amplitude and
phase as a function of NP position in the $(x,y)$ plane at different axial
positions $z$. Linear grey scale from $-\pi$ to $\pi$ for phases,
and from $m$ to $M$ for field amplitudes. The amplitude ratio is on
a logarithmic scale over 4 orders of magnitudes. Experiment: Top
inset shows a sketch of the sample and a TEM image of a typical NP
from the batch used. Pump (probe) power at the sample was 18\,$\mu$W
(9\,$\mu$W), 3\,ms pixel dwell time,
0.5\,ps pump-probe delay time, pixel size in-plane (axial) was
17\,nm (75\,nm). Calculations assume particle asphericity in plane with semi-axes $a=30.135$\,nm and
$b=30$\,nm (see text). \label{figexp1}}
\end{figure*}

The experimental data shown in \Fig{figexp1} are
scans of the $(x,y,z)$ sample position. They reveal all main features seen in the calculations, namely
a ring-like spatial distribution in-plane of the amplitude ratio
$\AF^- / \AF^+$ and a phase of the cross-polarized component, and in
turn of the FWM ratio, twice rotating from 0 to $2\pi$ along the
in-plane polar angle. A main difference to the calculations is the
observation of two displaced nodes, rather than a single central
node, in the amplitude of the cross-polarized component, and in turn
a nearly constant phase of the cross-polarized component in the
central area. The FWM ratio resolved across different axial planes
is also shown in \Fig{figexp1}. We observe that, besides these
minima, the axially resolved distributions agree with the calculated
behavior of a linear relationship between $\PhiF^+$ and $z$ and a
$z$-dependent $\Theta_0$ manifesting as a rotated phase pattern of
$\Theta=\PhiF^- - \PhiF^+$ for different $z$-planes.

Notably, by measuring on $>50$ particles in the sample we could not
find any pattern forming a $l=2$ optical vortex as simulated in
\Fig{figSim}. Different particles showed minima with different
relative displacement, or single displaced minima, or no minima
within the spatial range of sufficient signal to noise (see
Supplementary Information Fig.\,S12), suggesting that these minima
are related to physical differences between particles. Indeed, we
can explain these experimental findings assuming a small particle
asphericity. This is shown in the calculations in
Fig.\,\ref{figexp1} where we used an ellipsoid nanoparticle with
semi-axes $a=30.135$\,nm and $b=c=30$\,nm along the $x$, $y$ and $z$
axis in the particle reference system, which is rotated in plane
relative to the laboratory system (see also see Supplementary
Information Fig.\,S5 and S6). It is remarkable that an asymmetry of
only 0.5\% in aspect ratio, or about one atomic layer of gold,
manifests as a significant perturbation of the cross-polarized field
patterns compared to the spherical case. With such sensitivity to
asymmetry, the lack of experimentally observed pattern corresponding
to a perfectly spherical particle is not surprising, considering the
real particle morphology as observed in transmission electron
microscopy (TEM) for this sample (see top inset in \Fig{figexp1} and
Supplementary Information Fig.\,S8).

\begin{figure*}[t!]
\includegraphics*[width=16cm]{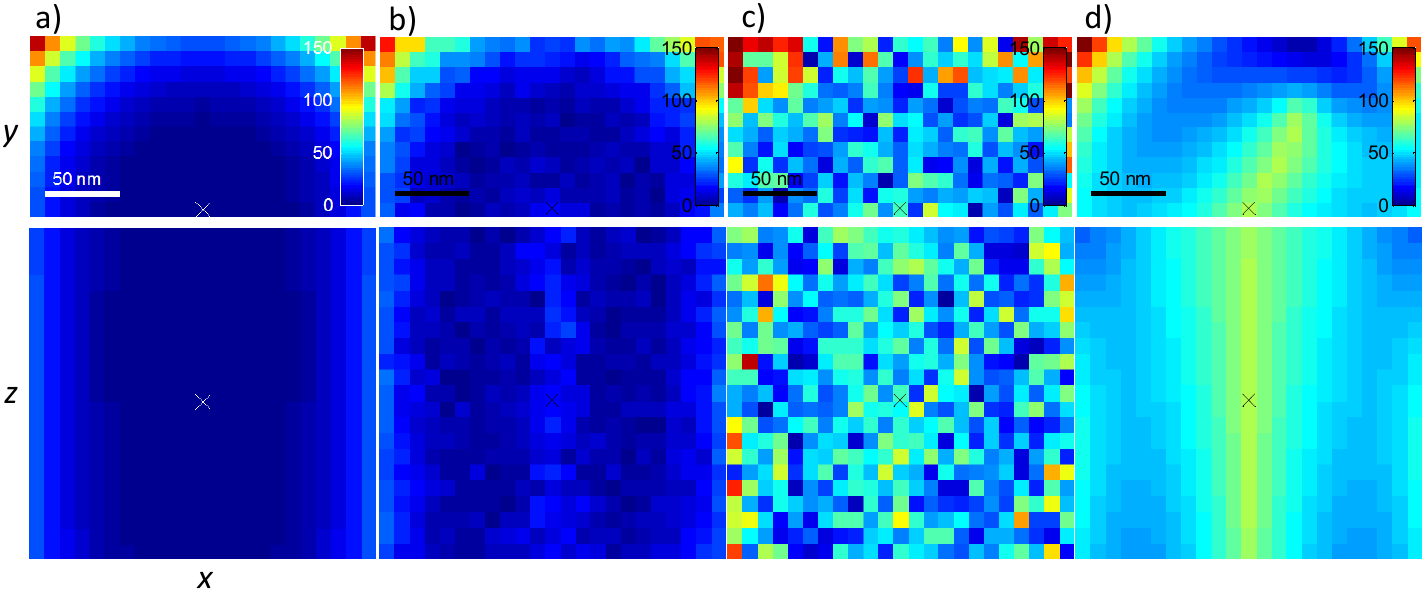}
\caption{{\bf Position localisation accuracy.} Difference between
the set position of the nanoparticle in 3D and the deduced position
using the FWM field amplitude and phase with the coordinate
reconstruction parameters as described in the text. The color bar
gives the values of this difference in nanometer. Top panels show
maps in the $(x,y)$ plane, bottom panels are axial $(x,z)$ maps. The
cross shows the position of the focus center. a) Spherical particle
without shot-noise; b) Spherical particle with relative shot-noise
of 0.07\% as in the experiment in \Fig{figexp1}; c) spherical
particle with relative shot-noise of 0.7\%; d) asymmetric particle
with semi-axis $a=30.135$\,nm and $b=c=30$\,nm, as in the
simulations in \Fig{figexp1}. \label{figaccuracy}}
\end{figure*}

The shot-noise in the experiment as well as the deviation from
perfect sphericity affect the localisation accuracy. This is
analyzed in \Fig{figaccuracy}. Firstly, we performed simulations as
in \Fig{figSim} for a perfectly spherical particle including
experimental shot-noise. The experimental shot-noise
was evaluated by taking the statistical distribution of the measured
FWM field (in both the in-phase $Re$ and in-quadrature $Im$ components) in a spatial region away
from the particle, where no FWM is detectable. The standard
deviation $\sigma$ of this distribution was deduced, and was found
to be identical in both components, and for the
co-polarised and cross-polarised components, as expected for an
experimental noise dominated by the shot-noise in the reference beam
(see Supplementary Information Fig.\,S9 for the dependence of
$\sigma$ on the power in the reference beam). A relative noise
figure was defined as $\sigma/A_0$ with $A_0$ being the maximum
measured value of the co-polarized FWM field amplitude. The
simulations were performed using a statistical distribution of the
FWM field values ($Re$ and $Im$ components) at each spatial pixel
having the same relative noise $\sigma/A_0$ as the experimental
data.

To quantify the uncertainty in the localisation of the particle
coordinates, we then calculated the deviation
$\Delta=\sqrt{(x-x_{\rm s})^2+(y-y_{\rm s})^2+(z-z_{\rm s})^2}$
between the set position of the NP in 3D ($x_{\rm s},y_{\rm
s},z_{\rm s}$) and the deduced position ($x,y,z$) using the FWM
field amplitude and phase as detailed above. The resulting $\Delta$
in the focal plane $(x,y,0)$ and in the $(x,0,z)$ cross-section
along the axial direction are shown in \Fig{figaccuracy}a for the
ideal case as in Fig.\ref{figSim}, and in \Fig{figaccuracy}b when
adding a relative noise of 0.07\% corresponding to the experiment in
\Fig{figexp1}. $\Delta\neq0$ in the ideal case is a measure of the
validity of the assumed dependencies, in particular the quadratic
behavior in the radial coordinate $R$, which becomes inaccurate for
$R>60$\,nm as discussed. This deviation can be easily removed by
including a $R^4$ term (see Supplementary Information Fig.\,S3).
Notably, in the region where the assumed trends are valid, we
observe that adding the experimental noise results in a localization
uncertainty of less than 20\,nm. Increasing the relative noise by
one order of magnitude to 0.7\% results in a localization
uncertainty of $\sim50$\,nm, as shown in \Fig{figaccuracy}c. We
emphasize that $\Delta$ is dominated by the in-plane uncertainty,
i.e. $\Delta=\sqrt{(x-x_{\rm s})^2+(y-y_{\rm s})^2+(z-z_{\rm
s})^2}\cong\sqrt{(x-x_{\rm s})^2+(y-y_{\rm s})^2}$. In fact,
considering that the axial direction $z$ is determined directly by
the slope of $\PhiF^+$ and does not involve the cross-polarized FWM,
we find $z$ localization accuracies as small as 0.3\,nm for 0.7\% noise
in $\PhiF^+$.

Note that the relative shot-noise in the experiments scales as
$1/\left(V\sqrt{t}{I_1}\sqrt{I_2}\right)$ where $V$ is the particle
volume (in the Rayleigh regime), $I_1$ ($I_2$) is the intensity of
the pump (probe) beam at the sample, and $t$ is the acquisition
time. It can therefore be adjusted according to the requirements.
For example, a 15\,nm radius NP imaged under the experimental
conditions as in \Fig{figexp1} would give rise to an 8-fold increase
in relative noise, hence still maintaining the localization
uncertainty to below 50\,nm. These results elucidate that a
localization accuracy of below 20\,nm in-plane and below 1\,nm in
axial direction is achievable with the proposed method with a
realistic shot-noise level as in the experiment.

However, the lack of particle symmetry is a limitation for the
in-plane localization. Introducing the particle asymmetry, without
shot-noise, as calculated in \Fig{figexp1}, results in a significant
deviation $\Delta\sim100$\,nm in the central area, due to the lack
of a central node in the cross-polarized FWM amplitude (see
\Fig{figaccuracy}d). On the other hand, it is remarkable how
sensitive the described FWM technique is to particle asymmetry,
which can be used as a new tool to detect particle ellipticity
down to $a/b-1=10^{-4}$ (corresponding to atomic accuracy comparable
to TEM) as well as particle orientation. We find that the FWM amplitude ratio in the focus center scales linearly
with the particle ellipticity and that the phase
of the FWM ratio in the focus center scales with the in-plane
particle orientation angle (which for the data in Fig.\ref{figexp1} was found to be $\gamma=150^\circ$, see Supplementary Information Fig.\,S5
and S6).

The limitation of particle asymmetry can be overcome by improving
colloidal fabrication techniques. For example, it has been reported
in the literature that gold nanorods can be synthesized as single
crystals without stacking faults or
dislocations\,\cite{KatzBonnNL11}, and can be shaped to become
spherical particles under high-power femtosecond laser
irradiation\,\cite{TaylorACSNano14}. We also note that when
examining gold NPs of 5\,nm radius (see Supplementary Information
Fig.\,S8 and S13), we found a large proportion ($\sim70\%$) having a
FWM amplitude ratio $\AF^-/\AF^+ \leq0.02$ in the center of the
focal plane, in this case limited by the signal-to-noise ratio
rather than the particle asymmetry, suggesting that smaller NPs
might be intrinsically more mono-crystalline hence symmetric.

Importantly, when NPs are not immobilized onto a surface but freely
rotating (a more relevant scenario for particle tracking
applications) we can expect that the rotational averaging over the
acquisition time (the rotational diffusion constant in water at room
temperature of a 30\,nm radius nanoparticle is
$\sim10^4$\,rad${^2}$/s) would result in an effective symmetry, and
would lead to a localization accuracy only limited by shot-noise.
This case is discussed in the next section.

\section{Rotational averaging}

\begin{figure*}[t!]
\includegraphics*[height=12cm]{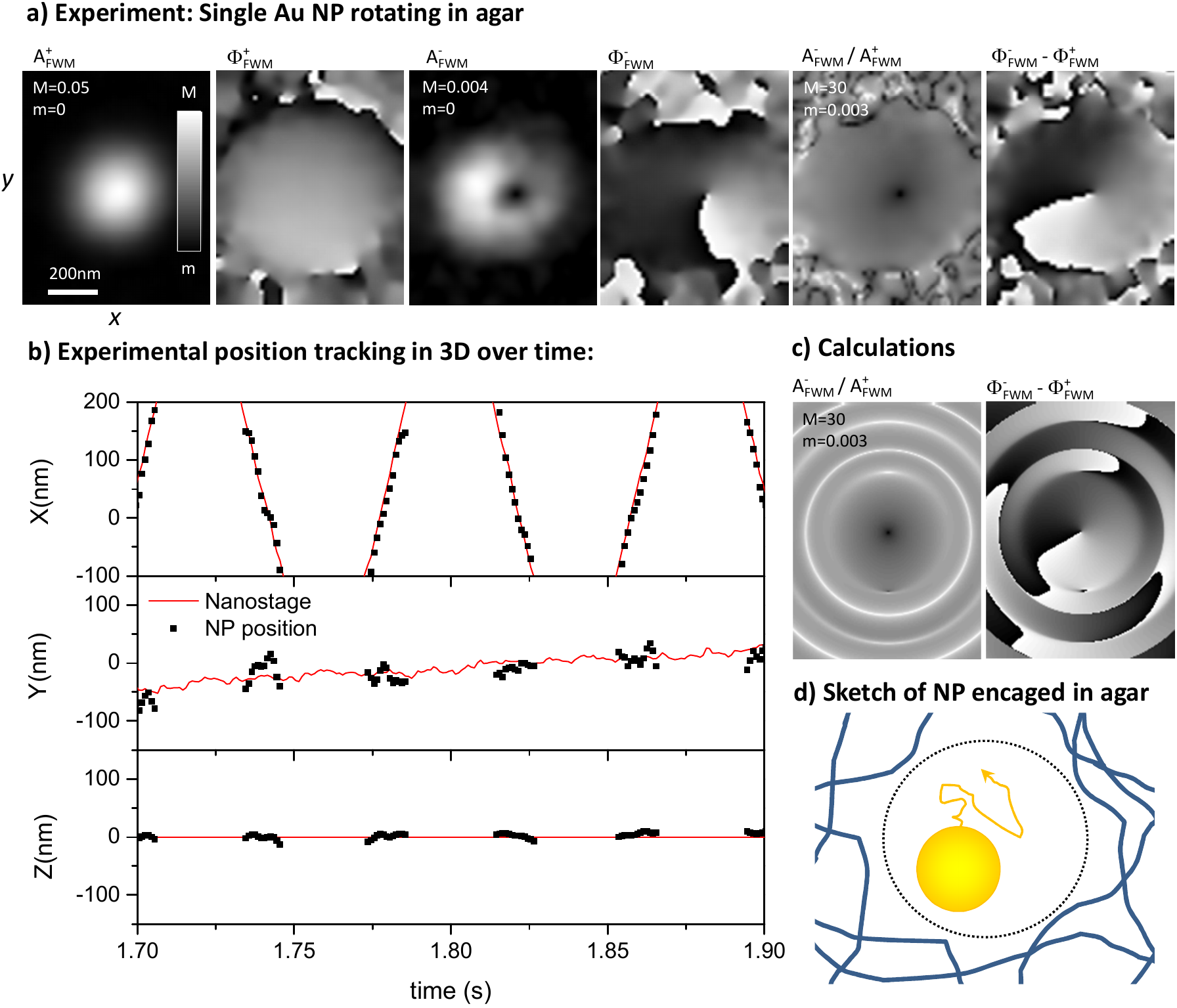}
\caption{{\bf Rotational averaging of nanoparticle asymmetry.}
Single 25\,nm radius gold NP freely rotating while encaged in an
agarose gel pocket (see sketch). a) Experimental $xy$ scan of the
cross and co-circularly polarised FWM field in the focal plane,
using a 1.27\,NA water-immersion objective. Linear grey scale from
$-\pi$ to $\pi$ for phases, and from $m$ to $M$ for field amplitudes
with $M$ given relative to the maximum $\Ar^+$. The amplitude ratio
is on a logarithmic scale over 4 orders of magnitudes. Pump (probe)
power at the sample was 70\,${\mu}$W (10\,${\mu}$W) with pump (probe) filling factor 2.15 (0.97). Measurements
were performed with 0.5\,ms pixel dwell time, 0.5\,ps pump-probe
delay time, and 13\,nm pixel size in-plane. Data are shown as
spatial averages over an effective area of $3\times3$ pixels. b)
Time traces of the retrieved particle position coordinates in 3D
from the measured FWM amplitude and phase (symbols) compared with
the coordinates recorded from the scanning piezoelectric sample
stage (lines). Traces are binned to an equivalent 1\,ms acquisition
time. c) Calculated FWM field ratio assuming a polarisation tensor
that projects the longitudinally polarised field component into the
$(x,y)$ plane (see text). \label{figfreerot}}
\end{figure*}

To mimic a relevant biological environment such as the cytosolic
network, as well as having single NPs freely rotating but not
diffusing out of focus, gold NPs of 25\,nm radius were embedded in a
dense (5\% w/v) agarose gel in water (see Supplementary Information
section S2.ii) and measured using an index-matched 1.27\,NA
water-immersion objective. For these experiments the focused size of
the pump beam was increased by a factor of two (by under-filling the back focal aperture), to enlarge the
region where a single NP could diffuse while still being excited by
the pump field hence giving rise to FWM (this also increases the maximum cross-polarized FWM amplitude). Conversely, the probe beam
was tightly focused to exploit the full NA of the objective an in
turn exhibit an $l=2$ optical vortex in the cross-circularly
polarised component as calculated in Fig.\,\ref{figSim}.

Fig.\,\ref{figfreerot}a shows an $xy$ scan of the measured cross and
co-circularly polarised FWM field in the focal plane, and the
corresponding ratio in amplitude and phase, on a single NP while
freely rotating but being enclosed in a tight pocket of the agar
network resulting in a negligible average translation during the
measurement time (for an estimate of the size of the pocket see
Supplementary Information section S2.xi). As expected, the
rotational averaging enables the observation of an optical vortex
with a central amplitude node in the cross-circulalry polarised FWM.
Surprisingly however, the phase pattern reveals an $l=1$ vortex, as
opposed to the predicted $l=2$, i.e. the phase is directly changing
with the in-plane polar angle instead of twice. Calculations show
that for an incident circularly polarised field, the longitudinally
polarised component ($E_{\rm z}$) in the focus of a high NA
objective has this symmetry, and we can reproduce the experimental
findings by introducing a particle polarisability tensor which
projects the longitudinal component into the $xy$ plane (see
Supplementary Information Section S1.vii). Conversely, it is not
possible to reproduce the experimental findings by assuming a
non-rotating randomly-oriented asymmetric particle (see
Supplementary Information Section S2.xii), hence free rotation is
key to the pattern observed in Fig.\,\ref{figfreerot}a. Since the
only particle asymmetry that does not average upon rotation is 3D
chirality, we suggest that this response is a manifestation of
chirality (albeit a detailed theoretical understanding of the
nonlinear optical response of a chiral particle is beyond the scope
of this work). We remind that these quasi-spherical NPs have
irregularities of few atoms clusters which can lead to symmetry
breaking, as already seen for their ellipticity, and thus also 3D
chirality (see TEM in Supplementary Information Fig.\,S8). Notably,
an $l=1$ optical vortex offers a simpler dependence for the
retrieval of the NP position coordinates, with a FWM amplitude ratio
$\AF^-/\AF^+$ scaling linearly with $R$ and a phase $\Theta=\PhiF^-
- \PhiF^+$ directly changing with the polar angle $\varphi$ (see
Supplementary Information Fig.\,S14). Using these dependencies
(alongside the axial position $z$ given by $\PhiF^+$ as previously
shown), we have retrieved the NP position coordinates in 3D for the
scan shown in Fig.\,\ref{figfreerot}a and compared them with the
position coordinates recorded from the scanning piezoelectric sample
stage. This is shown in Fig.\,\ref{figfreerot}b. We observe
deviations below 50\,nm in $x,y$ between the retrieved NP positions
and the coordinates from the stage and a remarkable accuracy in $z$
better than 10\,nm, limited by systematic drifts rather than
shot-noise. From the shot noise $\sigma/A_0=0.007$ in these
measurements and the coordinate retrieval parameters, we deduce the
uncertainty to be below 20\,nm in $x,y$ and 1\,nm in $z$, using only
1\,ms acquisition time.

\section{Single particle tracking}

\begin{figure*}[t!]
\includegraphics*[height=16.8cm]{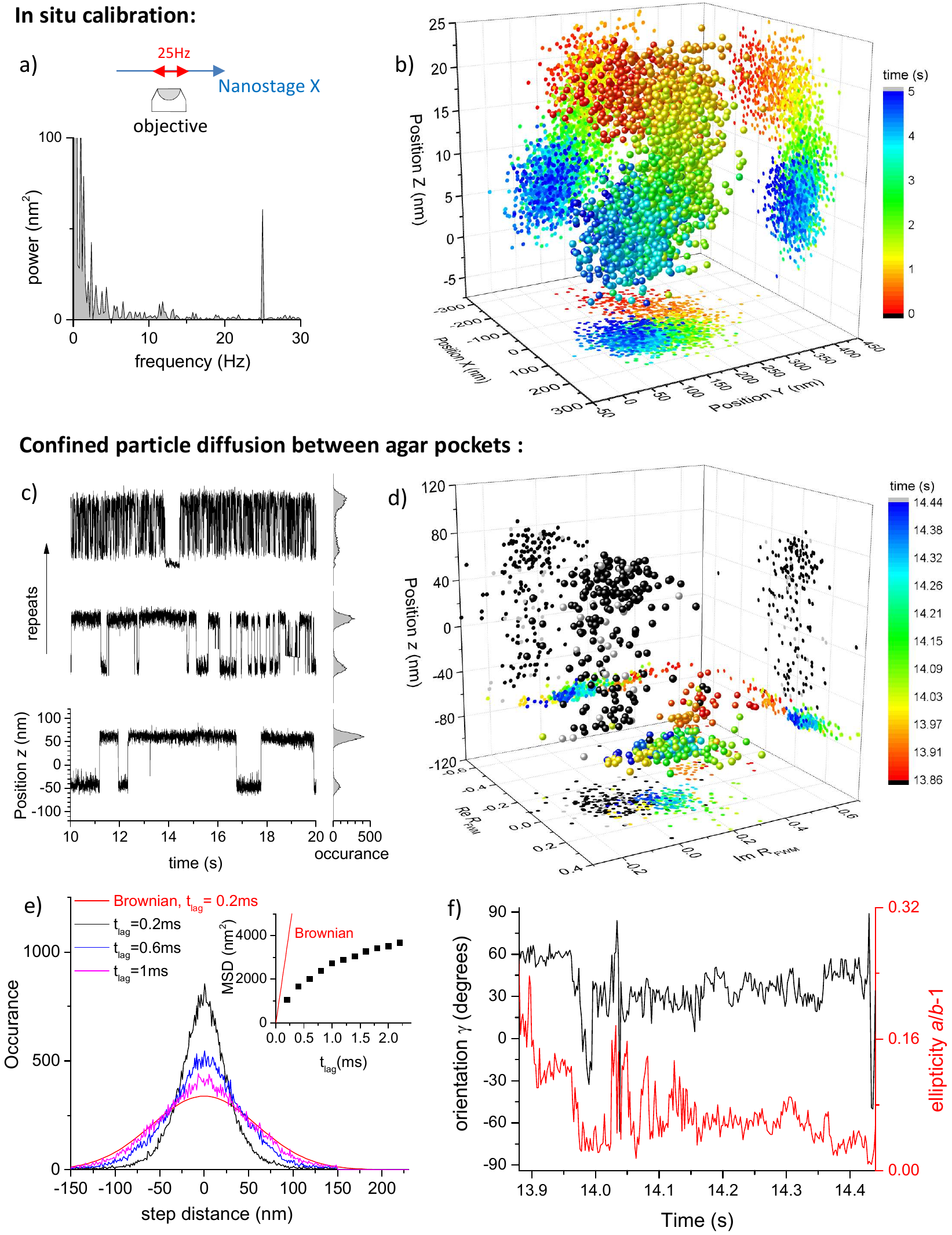}
\caption{{\bf In situ calibration and nanoparticle tracking.} a)
Power spectrum of FWM field ratio showing the 25\,Hz oscillation
imposed onto the sample stage for calibration, as sketched. b)
Retrieved position coordinates versus time of a single 25\,nm radius
NP in agar (2\,ms per point). c) Axial position
time traces of a single gold NP while jumping between two gel
pockets. d) Zoom of top trace in (c) showing the axial position and FWM field
ratio dominated by the NP asymmetry; corresponding in-plane ellipticity and
orientation are in f). e) Analysis of axial position time traces
acquired at 0.2\,ms per point (see text). \label{figtracking}}
\end{figure*}

A demonstration of the practical applicability of the method for
scanless single particle tracking in 3D is shown in
Fig.\,\ref{figtracking} on gold NPs of 25\,nm radius embedded in
a dense (5\% w/v) agarose gel in water, which provides a
heterogeneous environment as shown in Fig.\,\ref{figcell}(bottom).
For practical purposes, we implemented a simple {\it
in situ} calibration of the in-plane NP position coordinates without
the need for prior knowledge and/or characterization of the particle
optical response. As shown in Fig.\,\ref{figtracking}a, we apply a small
oscillation of known amplitude (16\,nm) and frequency (25\,Hz) to
one axis of the sample stage. When the FWM field ratio from the NP
encodes the NP in-plane position as discussed in the previous
section, this oscillation is detected in the measured $\EF^- /
\EF^+$ with an amplitude and a phase which we can use to directly
calibrate the FWM field ratio in terms of in-plane position
coordinates of known size and direction (see also Supplementary
Information section S2 x.iii). Fig.\,\ref{figtracking}a shows the
power spectrum of the Fourier transform of $\EF^- / \EF^+$ which
exhibits a peak at this modulation frequency. The correspondingly calibrated
particle position coordinates over time are shown
Fig.\,\ref{figtracking}b. Conversely, if for a NP we do not observe this
oscillation in $\EF^- / \EF^+$, the FWM field
ratio is a measure of the NP in-plane asymmetry and orientation. This case is shown in
Fig.\,\ref{figtracking}c-d (for the power spectrum see Supplementary
Information section S2 x.iii). Importantly, we can still accurately
measure the axial position coordinate, which is directly given by
$\PhiF^+$ and does not involve the cross-polarized FWM, while now
$\EF^- / \EF^+$ encodes information on the NP asymmetry and
orientation, rather than its in-plane position.
Fig.\,\ref{figtracking}c shows examples of traces of the NP $z$
position acquired over several tens of seconds (with 2\,ms point
acquisition) from which we can see a 'jumping behavior' around two
preferred axial locations, suggesting the presence of two pockets in
the agar gel separated by about 100\,nm. Fig.\,\ref{figtracking}d shows the $z$ position
together with $\EF^- / \EF^+$ (as real and imaginary parts) as a
zoom over a time window (from the top trace in (c)) during which the NP
got stuck in a corner, below the center of the lower pocket. From the strong and slowly varying FWM field ratio, the corresponding time evolution of the NP
in-plane orientation angle and aspect ratio is given in
Fig.\,\ref{figtracking}f. Finally, Fig.\,\ref{figtracking}e shows
the analysis from an axial position time trace acquired at high
speed (0.2\,ms time per point, for over 10\,s), indicating
constrained diffusion. The traces were analyzed by calculating the mean square displacement (MSD) as a
function of time lag ($t_{\rm lag}$)\,\cite{ManzoRPP15}. Histograms
of the displacement for different $t_{\rm lag}$ are shown in
Fig.\,\ref{figtracking}e. The variance of the histogram gives the
MSD, which is plotted versus $t_{\rm lag}$ in the inset. For
Brownian diffusion MSD$=2Dt_{\rm lag}$ (for each dimension in
space), with the free diffusion coefficient $D$ in water given in
the Supplementary Information section S2 xi. The resulting linear behavior is
shown as red line (and red Gaussian histogram) in
Fig.\,\ref{figtracking}e. The observed sub-linear dependence and
saturation of MSD versus $t_{\rm lag}$ indicates confined diffusion.

%\end{section}

\begin{section}{Discussion and Conclusion}
We have shown a new method to determine the position of a single
non-fluorescing gold NP with nanometric accuracy in three dimensions
from scanless far-field optical measurements. The method is based on
the interferometric detection of the polarisation-resolved resonant
FWM field in amplitude and phase, with a high numerical aperture
objective. The displacement of the nanoparticle from the center
focus in the axial direction is directly determined from the
epi-detected FWM field phase, while the in-plane displacement
manifests as a cross-circularly polarized component due to the
optical vortex field pattern in the focus of a high numerical
aperture objective, the amplitude and phase of which enables
accurate position retrieval.

Shape asymmetry of the NP of as little as 0.5\% ellipticity,
corresponding to about one atomic layer of gold, also induces a
cross-circularly polarized field. This exceptional sensitivity to
asymmetry eventually limits accurate position retrieval in-plane.
This can be overcome by observing a NP freely rotating, such that
rotational averaging of the asymmetry occurs over the acquisition
time.

Experimentally, we show that the FWM detection is completely
background-free in scattering environments such as biological
cells, and outperforms existing methods such as reflectometry,
scattering and differential interference contrast. A localization
accuracy of below 50\,nm in-plane, and below 10\,nm axial, limited by systematic drifts, is shown
with a single NP of 25\,nm  radius freely rotating in an agarose
gel, at an acquisition time of only 1\,ms. The shot-noise limited accuracy is found to be below 20\,nm in plane and 1\,nm axially. Smaller NPs can be
measured by correspondingly increasing the intensity of the incident
beams and/or the acquisition time. Notably, we find that, during
free rotation, the cross-circularly polarized FWM field distribution
in the focal plane is an $l=1$ optical vortex, as opposed to the
predicted $l=2$ for a perfectly spherical particle, and we suggest
that this is a manifestation of NP chirality not averaged during
rotation.

To demonstrate the practical applicability of the method for single
particle tracking, we provide two examples of NPs diffusing in a
dense agarose gel network. In one case, we show particle position
coordinates retrieved using an {\it in situ} calibration procedure
and the corresponding tracking in 3D. In the second case, we show
that when the cross-circularly polarized component is dominated by
the effect of shape asymmetry, it can be used for tracking the
in-plane asymmetry and orientation, while the axial position
coordinate provides information on the particle diffusion.

Ultimately, this method paves the way towards a new form of single
particle tracking, where not only the NP position, but also its
asymmetry, orientation, and chirality are detected with sub-millisecond time resolution, revealing much more
information about the NP and its complex dynamics (e.g. hindered
rotation) while moving and interacting within a disordered 3D
environment.

\end{section}

\section{Appendix: Methods}

{\bf FWM experimental set-up.} Optical pulses of 150\,fs duration
centered at 550\,nm wavelength with $\nu_{\rm L}=80$\,MHz repetition
rate were provided by the signal output of an optical parametric
oscillator (Newport/Spectra Physics, OPO Inspire HF 100) pumped by a
frequency-doubled femtosecond Ti:Sa laser (Newport/Spectra Physics,
Mai Tai HP). In the experiment, the amplitude modulation frequency
of the pump beam was $\nu_{\rm m}=0.4$\,MHz. The MO was either a
$100\times$ magnification oil-immersion objective of 1.45\,NA (Nikon
CFI Plan Apochromat lambda series) or a $60\times$ magnification
water-immersion objective of 1.27\,NA (Nikon CFI Plan Apochromat
lambda series, super resolution) mounted into a commercial inverted
microscope stand (Nikon Ti-U). The sample was positioned with
respect to the focal volume of the objective by an $xyz$
piezoelectric stage with nanometric position accuracy (MadCityLabs
NanoLP200). A prism compressor was used to pre-compensate the chirp
introduced by all the optics in the beam path, to achieve Fourier
limited pulses of 150\,fs duration at the sample. Furthermore, since
the reference beam does not travel through the microscope objective,
glass blocks of known group velocity dispersion were added to the
reference beam path, in order to match the chirp introduced by the
microscope optics and thus maximize the interference between the
reflected FWM field and the reference field at the detector. AOM$_2$ was driven at
$\nu_{2}=$82\,MHz. BS$_1$ is a 80:20 transmission:reflection
splitter, transmitting most of the signal. We used balanced silicon photodiodes (Hamamatsu S5973-02)
with home-built electronics and a high-frequency digital lock-in
amplifier (Z\"urich Instruments HF2LI) providing six dual-phase
demodulators, enabling to detect for both polarizations the carrier
at $\nu_2-\nu_{\rm L}=2$\,MHz and both sidebands at
$\nu_2\pm\nu_{\rm m}-\nu_{\rm  L}=2\pm0.4$\,MHz. This scheme
overcomes the limitation in our previous work of using two separated
lock-ins with associated relative phase offset\,\cite{MasiaPRB12}
and provides an intrinsic phase referencing and a noise reduction
via the detection of both side bands (see Supplementary Information
section S2.iv).

{\bf Numerical simulations.} The field in the focal area is
calculated using "PSF Lab" - a software package that allows
calculation of the point spread function of an aplanatic optical
system\,\cite{Nasse2010}. For the calculations in Fig.\,\ref{figSim}
and \Fig{figexp1} the simulation parameters - wavelength $\lambda$,
objective lens NA, coverslip thickness $d$, medium refractive index
$n$, back objective filling factor $\beta$ - were chosen to match
the actual experimental conditions in \Fig{figexp1}, namely
$\lambda=550$\,nm, NA=1.45, $d=0.17$\,mm, $n=1.5185$, and
$\beta=0.83$. The filling factor is defined as $\beta = a/w$, where
$a$ is the aperture radius of the objective lens and $w$ is the Gaussian parameter in the electric field radial
dependence at the objective aperture $E = E_{0} e^{(-r^2/w^2)}$. For
the calculations in Fig.\,\ref{figfreerot} the simulation parameters
were NA=1.27, $n=1.333$, $\beta=2.15$ for the pump and
$\beta=0.97$ for the probe beam. The calculation was performed for a
linear polarization of the incident field along the $x$ axis. This
results in a vectorial field at the focus called $\bE^x(\br)$. For
the orthogonal linear incident polarization, the results were
rotated counterclockwise to obtain $\bE^y$. To simulate circular incident
polarization, the calculated field maps $\bE^x$ and $\bE^y$ were
combined with complex coefficients, namely:
$\bE^+=\frac{1}{\sqrt{2}}\left(\bE^x+i\bE^y\right)$,
$\bE^-=\frac{1}{\sqrt{2}}\left(\bE^x-i\bE^y\right)$ for left and
right circular polarization, respectively.

%
%{\bf References go here}
%Create the reference section using BibTeX:
%\bibliography{fwm,langsrv,zoriniants}

\begin{thebibliography}{10}
\expandafter\ifx\csname url\endcsname\relax
  \def\url#1{\texttt{#1}}\fi
\expandafter\ifx\csname urlprefix\endcsname\relax\def\urlprefix{URL
}\fi \providecommand{\bibinfo}[2]{#2}
\providecommand{\eprint}[2][]{\url{#2}}

\bibitem{BetzigScience92}
\bibinfo{author}{Betzig, E.} \& \bibinfo{author}{Trautman, J.~K.}
\newblock \bibinfo{title}{Near-field optics: Microscopy, spectroscopy, and
  surface modification beyond the diffraction limit}.
\newblock \emph{\bibinfo{journal}{Science}} \textbf{\bibinfo{volume}{257}},
  \bibinfo{pages}{189--195} (\bibinfo{year}{1992}).

\bibitem{SanchezPRL99}
\bibinfo{author}{S\'{a}nchez, E.~J.}, \bibinfo{author}{Novotny, L.} \&
  \bibinfo{author}{Xie, X.~S.}
\newblock \bibinfo{title}{Near-field fluorescence microscopy based on
  two-photon excitation with metal tips}.
\newblock \emph{\bibinfo{journal}{Phys. Rev. Lett.}}
  \textbf{\bibinfo{volume}{82}}, \bibinfo{pages}{4014--4017}
  (\bibinfo{year}{1999}).

\bibitem{SchuckPRL05}
\bibinfo{author}{Schuck, P.~J.}, \bibinfo{author}{Fromm, D.~P.},
  \bibinfo{author}{Sundaramurthy, A.}, \bibinfo{author}{Kino, G.~S.} \&
  \bibinfo{author}{Moerner, W.~E.}
\newblock \bibinfo{title}{Improving the mismatch between light and nanoscale
  objects with gold bowtie nanoantennas}.
\newblock \emph{\bibinfo{journal}{Phys. Rev. Lett.}}
  \textbf{\bibinfo{volume}{94}}, \bibinfo{pages}{017402}
  (\bibinfo{year}{2005}).

\bibitem{RustNatMeth06}
\bibinfo{author}{Rust, M.~J.}, \bibinfo{author}{Bates, M.} \&
  \bibinfo{author}{Zhuang, X.}
\newblock \bibinfo{title}{Sub-diffraction-limit imaging by stochastic optical
  reconstruction microscopy (storm)}.
\newblock \emph{\bibinfo{journal}{Nature Methods}}
  \textbf{\bibinfo{volume}{3}}, \bibinfo{pages}{793--795}
  (\bibinfo{year}{2006}).

\bibitem{BetzigScience06}
\bibinfo{author}{Betzig, E.} \emph{et~al.}
\newblock \bibinfo{title}{Imaging intracellular fluorescent proteins at
  nanometer resolution}.
\newblock \emph{\bibinfo{journal}{Science}} \textbf{\bibinfo{volume}{313}},
  \bibinfo{pages}{1642--1645} (\bibinfo{year}{2006}).

\bibitem{HellScience07}
\bibinfo{author}{Hell, S.~W.}
\newblock \bibinfo{title}{Far-field optical nanoscopy}.
\newblock \emph{\bibinfo{journal}{Science}} \textbf{\bibinfo{volume}{316}},
  \bibinfo{pages}{1153--1158} (\bibinfo{year}{2007}).

\bibitem{FujiwaraJCB02}
\bibinfo{author}{Fujiwara, T.}, \bibinfo{author}{Ritchie, K.},
  \bibinfo{author}{Murakoshi, H.}, \bibinfo{author}{Jacobson, K.} \&
  \bibinfo{author}{Kusumi, A.}
\newblock \bibinfo{title}{Phospholipids undergo hop diffusion in
  compartmentalized cell membrane}.
\newblock \emph{\bibinfo{journal}{The Journal of Cell Biology}}
  \textbf{\bibinfo{volume}{157}}, \bibinfo{pages}{1071--1081}
  (\bibinfo{year}{2002}).

\bibitem{UenoBJ10}
\bibinfo{author}{Ueno, H.} \emph{et~al.}
\newblock \bibinfo{title}{Simple dark-field microscopy with nanometer spatial
  precision and microsecond temporal resolution}.
\newblock \emph{\bibinfo{journal}{Biophysical Journal}}
  \textbf{\bibinfo{volume}{98}}, \bibinfo{pages}{2014--2023}
  (\bibinfo{year}{2010}).

\bibitem{GuiAC12}
\bibinfo{author}{Gu, Y.}, \bibinfo{author}{Di, X.}, \bibinfo{author}{Sun, W.},
  \bibinfo{author}{Wang, G.} \& \bibinfo{author}{Fang, N.}
\newblock \bibinfo{title}{Three-dimensional super-localization and tracking of
  single gold nanoparticles in cells}.
\newblock \emph{\bibinfo{journal}{Anal. Chem.}} \textbf{\bibinfo{volume}{84}},
  \bibinfo{pages}{4111--4117} (\bibinfo{year}{2012}).

\bibitem{Ortega-ArroyoPCCP12}
\bibinfo{author}{Ortega-Arroyo, J.} \& \bibinfo{author}{Kukura, P.}
\newblock \bibinfo{title}{Interferometric scattering microscopy (iscat): new
  frontiers in ultrafast and ultrasensitive optical microscopy}.
\newblock \emph{\bibinfo{journal}{Phys. Chem. Chem. Phys.}}
  \textbf{\bibinfo{volume}{14}}, \bibinfo{pages}{15625--15636}
  (\bibinfo{year}{2012}).

\bibitem{LasneBJ06}
\bibinfo{author}{Lasne, D.} \emph{et~al.}
\newblock \bibinfo{title}{Single nanoparticle photothermal tracking (snapt) of
  5-nm gold beads in live cells}.
\newblock \emph{\bibinfo{journal}{Biophysical J.}}
  \textbf{\bibinfo{volume}{91}}, \bibinfo{pages}{4598--4604}
  (\bibinfo{year}{2006}).

\bibitem{LasneOE07}
\bibinfo{author}{Lasne, D.} \emph{et~al.}
\newblock \bibinfo{title}{Label-free optical imaging of mitochondria in live
  cells}.
\newblock \emph{\bibinfo{journal}{Opt. Express}} \textbf{\bibinfo{volume}{15}},
  \bibinfo{pages}{14184--14193} (\bibinfo{year}{2007}).
\newblock \bibinfo{note}{And references therein.}

\bibitem{MasiaOL09}
\bibinfo{author}{Masia, F.}, \bibinfo{author}{Langbein, W.},
  \bibinfo{author}{Watson, P.} \& \bibinfo{author}{Borri, P.}
\newblock \bibinfo{title}{Resonant four-wave mixing of gold nanoparticles for
  three-dimensional cell microscopy}.
\newblock \emph{\bibinfo{journal}{Opt. Lett.}} \textbf{\bibinfo{volume}{34}},
  \bibinfo{pages}{1816--1818} (\bibinfo{year}{2009}).

\bibitem{MasiaPRB12}
\bibinfo{author}{Masia, F.}, \bibinfo{author}{Langbein, W.} \&
  \bibinfo{author}{Borri, P.}
\newblock \bibinfo{title}{Measurement of the dynamics of plasmons inside
  individual gold nanoparticles using a femtosecond phase-resolved microscope}.
\newblock \emph{\bibinfo{journal}{Phys. Rev. B}} \textbf{\bibinfo{volume}{85}},
  \bibinfo{pages}{235403} (\bibinfo{year}{2012}).

\bibitem{bohren1998absorption}
\bibinfo{author}{Bohren, C.~F.} \& \bibinfo{author}{Huffman, D.~R.}
\newblock \emph{\bibinfo{title}{Absorption and Scattering of Light by Small
  Particles}} (\bibinfo{publisher}{Wiley-VCH}, \bibinfo{year}{1998}).

\bibitem{MilesACSPhotonics15}
\bibinfo{author}{Miles, B.~T.} \emph{et~al.}
\newblock \bibinfo{title}{Sensitivity of interferometric cross-polarization
  microscopy for nanoparticle detection in the near-infrared}.
\newblock \emph{\bibinfo{journal}{ACS Photonics}} \textbf{\bibinfo{volume}{2}},
  \bibinfo{pages}{1705--1711} (\bibinfo{year}{2015}).

\bibitem{KatzBonnNL11}
\bibinfo{author}{Katz-Boon, H.} \emph{et~al.}
\newblock \bibinfo{title}{Three-dimensional morphology and crystallography of
  gold nanorods}.
\newblock \emph{\bibinfo{journal}{Nano Letters}} \textbf{\bibinfo{volume}{11}},
  \bibinfo{pages}{273} (\bibinfo{year}{2011}).

\bibitem{TaylorACSNano14}
\bibinfo{author}{Taylor, A.~B.}, \bibinfo{author}{Siddiquee, A.~M.} \&
  \bibinfo{author}{Chon, J. W.~M.}
\newblock \bibinfo{title}{Below melting point photothermal reshaping of single
  gold nanorods driven by surface diffusion}.
\newblock \emph{\bibinfo{journal}{ACS Nano}} \textbf{\bibinfo{volume}{8}},
  \bibinfo{pages}{12071--12079} (\bibinfo{year}{2014}).

\bibitem{ManzoRPP15}
\bibinfo{author}{Manzo, C.} \& \bibinfo{author}{Garcia-Parajo, M.~F.}
\newblock \bibinfo{title}{A review of progress in single particle tracking:
  from methods to biophysical insights}.
\newblock \emph{\bibinfo{journal}{Rep. Prog. Phys.}}
  \textbf{\bibinfo{volume}{78}}, \bibinfo{pages}{124601}
  (\bibinfo{year}{2015}).

\bibitem{Nasse2010}
\bibinfo{author}{Nasse, M.~J.} \& \bibinfo{author}{Woehl, J.~C.}
\newblock \bibinfo{title}{Realistic modeling of the illumination point spread
  function in confocal scanning optical microscopy}.
\newblock \emph{\bibinfo{journal}{J. Opt. Soc. Am. A}}
  \textbf{\bibinfo{volume}{27}}, \bibinfo{pages}{295--302}
  (\bibinfo{year}{2010}).

\end{thebibliography}


\begin{section}{Acknowledgments}
This work was funded by the UK EPSRC Research Council (grant n.
EP/I005072/1, EP/I016260/1, EP/L001470/1, EP/J021334/1 and
EP/M028313/1) and by the EU (FP7 grant ITN-FINON 607842). P.B.
acknowledges the Royal Society for her Wolfson Research Merit award
(WM140077). We acknowledge Lukas Payne for help in sample
preparation and TEM characterization and Peter Watson, Paul Moody
and Arwyn Jones for help in cell culture protocols.
\end{section}
%\end{acknowledgments}

\begin{section}{Data availability}
Information about the data created during this research, including
how to access it, is available from Cardiff University data archive
at http://doi.org/10.17035/d.2017.0031438581.
\end{section}

\begin{section}{Author contributions}
P.B. and W.L. conceived the technique and designed the experiments.
W.L. wrote the acquisition software. F.M. performed preliminary
experiments and calculations. G.Z. performed the final calculations,
and the experiments on NPs attached onto glass. N.G. performed the
experiments on cells and prepared NPs in agar. W.L. and P.B.
performed the experiments on NPs in agar. P.B. wrote the manuscript.
All authors discussed and interpreted the results and commented on
the manuscript.
\end{section}

%%%%%%%%%% Merge with supplemental materials %%%%%%%%%%
\pagebreak
\widetext
\begin{center}
	\textbf{\large Background-free 3D nanometric localisation and sub-nm asymmetry detection of single plasmonic nanoparticles by four-wave mixing interferometry with optical vortices - Supplementary Information}
\end{center}
%%%%%%%%%% Merge with supplemental materials %%%%%%%%%%
%%%%%%%%%% Prefix a "S" to all equations, figures, tables and reset the counter %%%%%%%%%%
\setcounter{section}{0}
\setcounter{equation}{0}
\setcounter{figure}{0}
\setcounter{table}{0}
%\setcounter{page}{1}
%\makeatletter
\renewcommand{\theequation}{S\arabic{equation}}
\renewcommand{\thefigure}{S\arabic{figure}}
%\renewcommand{\bibnumfmt}[1]{[S#1]}
%\renewcommand{\citenumfont}[1]{S#1}
%%%%%%%%%% Prefix a "S" to all equations, figures, tables and reset the counter %%%%%%%%%%
\renewcommand{\thesection}{S\arabic{section}}
\renewcommand{\thefigure}{S\arabic{figure}}
\renewcommand{\thesubsection}{\roman{subsection}}

\section{Calculations}
\subsection{Polarisability of a non-spherical gold nanoparticle}

We describe a non-spherical nanoparticle (NP) as a metallic
ellipsoid with three orthogonal semi-axes of symmetry $a$, $b$ and
$c$. In the particle reference system the polarisability tensor
$\alphah$ is diagonal, and its eigenvalues are given
by\,\cite{bohren1998absorption}
\be \alpha_i= 4 \pi abc
\frac{\epsilon-\epsm}{3\epsm+3L_i(\epsilon-\epsm)}\,,\label{eqnalpha}\ee
where $\epsilon$ is the dielectric constant of the NP, $\epsm$ is the dielectric constant of the surrounding medium, and $L_i$ with $i=a,b,c$ are dimensionless quantities defined by the NP geometry as
\be\label{eq:L_i}\begin{aligned}
	L_a= \frac {abc}{2}\int_{0}^{\infty}
	(a^2+q)^{-\frac{3}{2}}(b^2+q)^{-\frac{1}{2}}(c^2+q)^{-\frac{1}{2}}d q \\
	L_b= \frac {abc}{2}\int_{0}^{\infty}
	(a^2+q)^{-\frac{1}{2}}(b^2+q)^{-\frac{3}{2}}(c^2+q)^{-\frac{1}{2}}d q \\
	L_c= \frac {abc}{2}\int_{0}^{\infty}
	(a^2+q)^{-\frac{1}{2}}(b^2+q)^{-\frac{1}{2}}(c^2+q)^{-\frac{3}{2}}d q
\end{aligned}
\ee
Only two out of three geometrical factors are independent, since for any ellipsoid $L_a+L_b+L_c=1$. For a spherical particle $a=b=c$, and
\be\label{eq:Laaa}L_1= L_2=L_3=\frac {a^3}{2}\int_{0}^{\infty}
(a^2+q)^{-5/2}d q =\frac{1}{3}\ee
For prolate spheroids (i.e. cigar-shaped) with $a>b=c$ or oblate
(i.e. pancake-shaped) spheroids with $a=b>c$ the geometrical factors
can be expressed analytically \cite{bohren1998absorption}. In the
most general case, we calculated the integrals in Eq.\,\ref{eq:L_i}
using the numerical integration function \verb"integral" of Matlab.
\begin{figure}[t!]
	\includegraphics*[width=16.5cm]{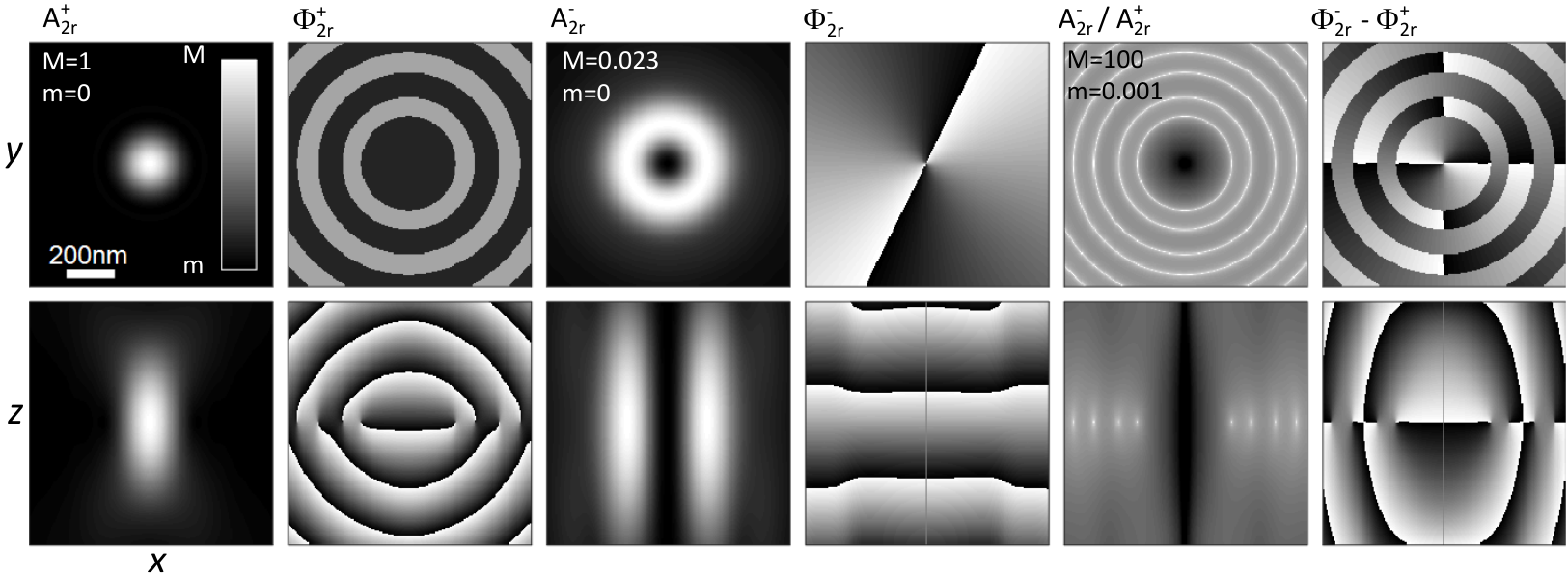}
	\caption{Calculated amplitude and phase
		of the reflected probe field $\Ar^\pm$ and $\Phir^\pm$, as function of the NP position, in the $(x,y)$ focal plane, and in the $(x,z)$ plane through the focus. Here $+$ refers to the co-polarised component and $-$ to the
		cross-polarised component relative to the left-circularly polarised
		incident probe. The calculation assumes a perfectly spherical gold
		NP in the dipole approximation. Linear grey scale from $-\pi$ to
		$\pi$ for phases, and from $m$ to $M$ for field amplitudes, as
		indicated. The amplitude ratio is shown on a logarithmic scale over
		5 orders of magnitude as indicated.} \label{sim}
\end{figure}

\subsection{Reflected probe field}

\Fig{sim} shows the reflected probe field as function of the NP position, calculated as in Fig.\,2 of the main manuscript, including its axial dependence through the focus.

\subsection{Pump-induced change of particle polarisability: Spatial distribution}

\begin{figure}[t!]
	\includegraphics*[width=6.5cm]{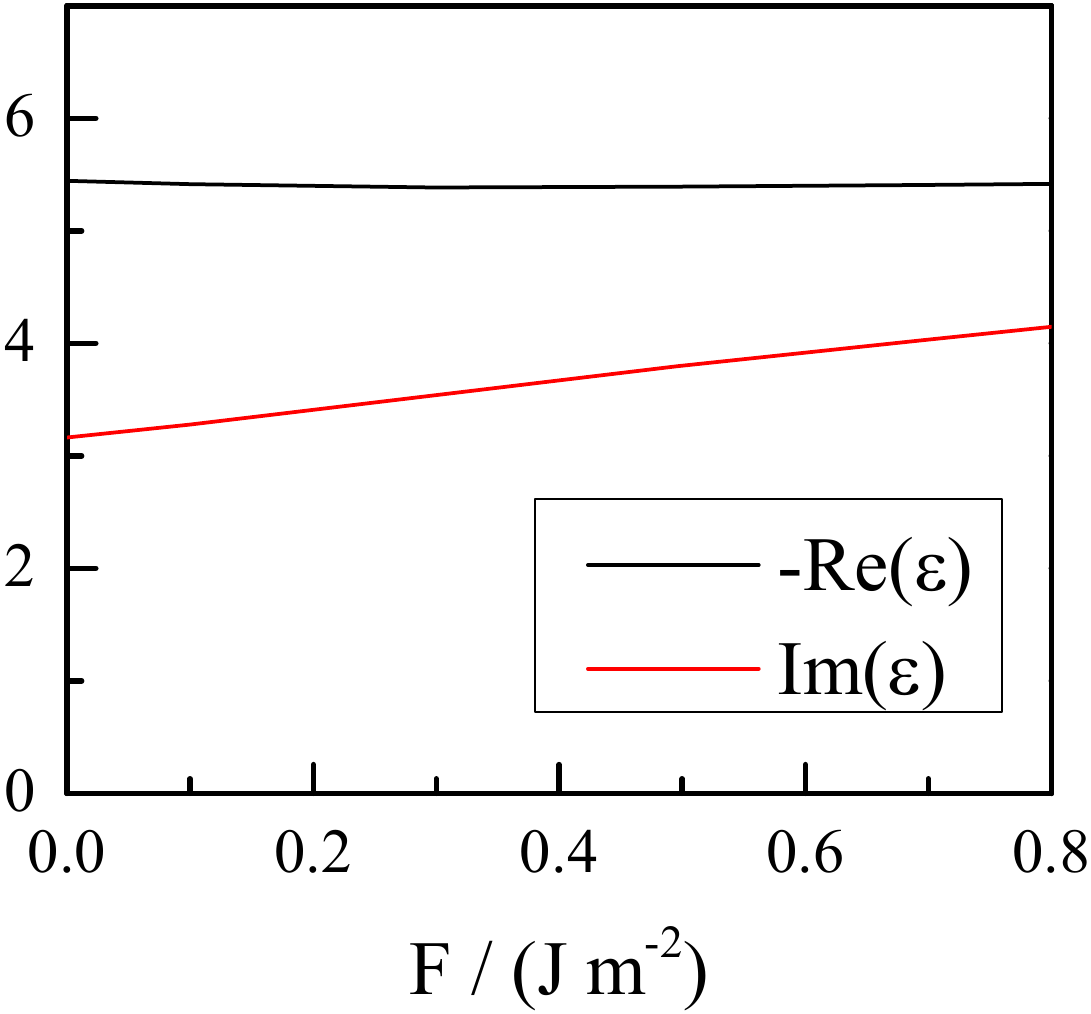}
	\caption{Calculated dependence of the real and imaginary part of the
		gold dielectric function $\epsilon$ assuming isotropic absorption by
		a 30\,nm radius gold NP as a function of the pump fluence, for a
		pump and probe wavelength of 550\,nm and a delay time of 0.5\,ps.
		Other parameters as described in the methods section of the main
		text.} \label{epsilon}
\end{figure}

To calculate the pump-induced change of the particle polarizability
$\delta\alphah$ we used the model developed in our previous
work\,\cite{MasiaPRB12}. Briefly, $\delta\alphah$ arises from the
transient change of the electron and lattice temperature following
the absorption of the pump pulse by the nanoparticle, and therefore
depends on the pump fluence at the NP, on the absorption
cross-section of the NP, and on the delay time between pump and
probe pulses. We calculated the change in the gold dielectric
function $\epsilon$ as in Ref.\,\onlinecite{MasiaPRB12} as a
function of pump fluence at the particle, using the absorption
cross-section of a 30\,nm radius gold NP, for a pump-probe
wavelength of 550\,nm and delay time of 0.5\,ps. The result is shown in
\Fig{epsilon}. To account for the spatial profile of the pump
we use the pump intensity spatial distribution $|\bE_1^-(\br)|^2$ to deduce $\epsilon(\br)$ in the presence of the pump, and in turn $\delta\alphah(\br)$.

\subsection{Parameters for NP position localisation}

\begin{figure*}[b!]
	\includegraphics*[width=16.5cm]{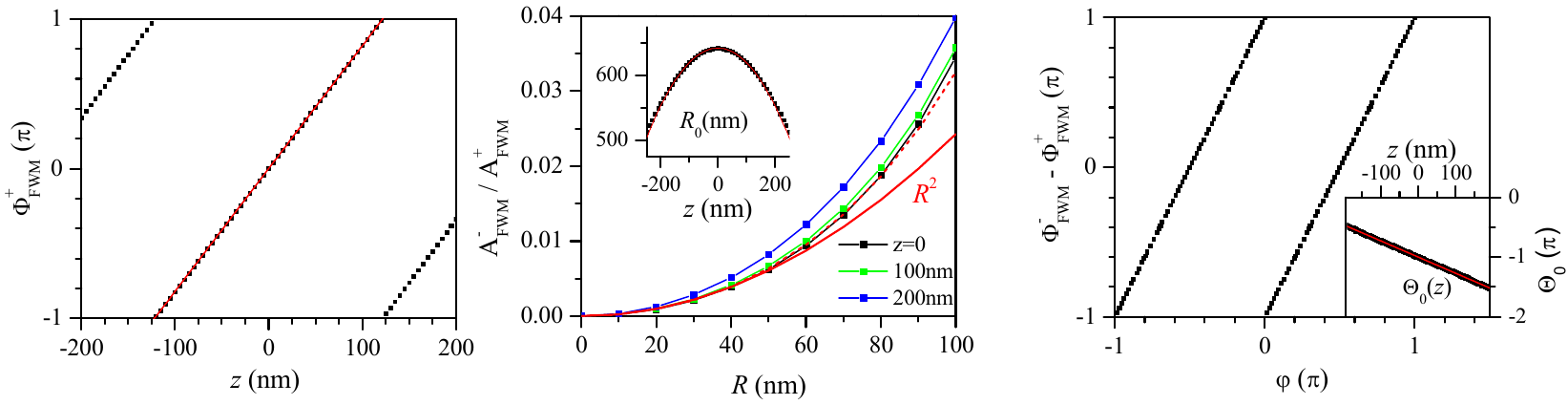}
	\caption{Left: Calculated axial dependence of the phase of the
		co-polarised FWM component (symbols) together with a linear fit (red
		line) having the slope $\partial z/\partial\Phi=38.8$\,nm/rad.
		Middle: Radial dependence of the cross to co-polarised FWM amplitude
		ratio for three different axial positions as indicated. The red
		solid line shows a fit assuming the quadratic dependence
		$(R/R_{0})^2$ at the focal plane $z=0$, while the red
		dashed line assumes the dependence
		$(R/R_{0})^2+(R/R_{1})^4$. Inset: Dependence of
		$R_{0}$ on the axial position. Right: Phase of the cross to
		co-polarised FWM  ratio $\Theta=\PhiF^- - \PhiF^+$ versus polar
		angle in the $z=0$ plane, following the relationship
		$\varphi=(\Theta-\Theta_0)/2$. Inset: $\Theta_0$ versus axial
		position. \label{figSim2}}
\end{figure*}

From the calculations in Fig.\,2 of the main manuscript we find a
linear relationship between $\PhiF^+$ and $z$ as shown in
\Fig{figSim2} with a slope $\partial z/\partial\Phi=38.8$\,nm/rad,
slightly larger than $\lambda/(4\pi n)=28.8$\,nm/rad. This is due to
the propagation of a focussed beam with high NA where a Gouy phase
shift occurs, effectively increasing the wavelength in axial
direction. Note that the measured $\PhiF^+$ has an offset, due to the phase of the reference field. In the shown representation, the value
of $\PhiF^+$ at the focal plane was taken as offset and subtracted,
such that \Fig{figSim2} directly gives $\partial z/\partial\Phi$.

For the in-plane radial coordinate $R$ of the NP position relative
to the focus position we find that the FWM amplitude ratio
$\AF^-/\AF^+$ scales quadratically with $R$ up to $R\sim60$\,nm,
such that this coordinate can be calculated as
$R=R_{0}\sqrt{\AF^-/\AF^+}$. $R_0$ slightly depends on the $z$
position as shown in Fig.\,\ref{figSim2}, a behavior which is well
described as $R_{0}(z)=R_{00}-Cz^2$ with $R_{00}=642.3$\,nm and
$C=0.002218$\,nm$^2$. Finally, the angular position coordinate $\varphi$
can be taken from the phase of the FWM ratio $\Theta=\PhiF^- -
\PhiF^+$ as shown in \Fig{figSim2}. We find
$\varphi=(\Theta-\Theta_0)/2$ with $\Theta_0=Bz-\pi$ and
$B=0.008514$\,rad/nm.

Note that there is a $\pi$ ambiguity in the angular position
coordinate $\varphi$ due to the scaling of the FWM ratio phase
proportional to $2\varphi$. This corresponds to an inversion of the
position in the $(x,y)$ plane.

\subsection{Dependence of $R_{0}$ on objective NA}

\begin{figure}[t!]
	\includegraphics*[width=7.5cm]{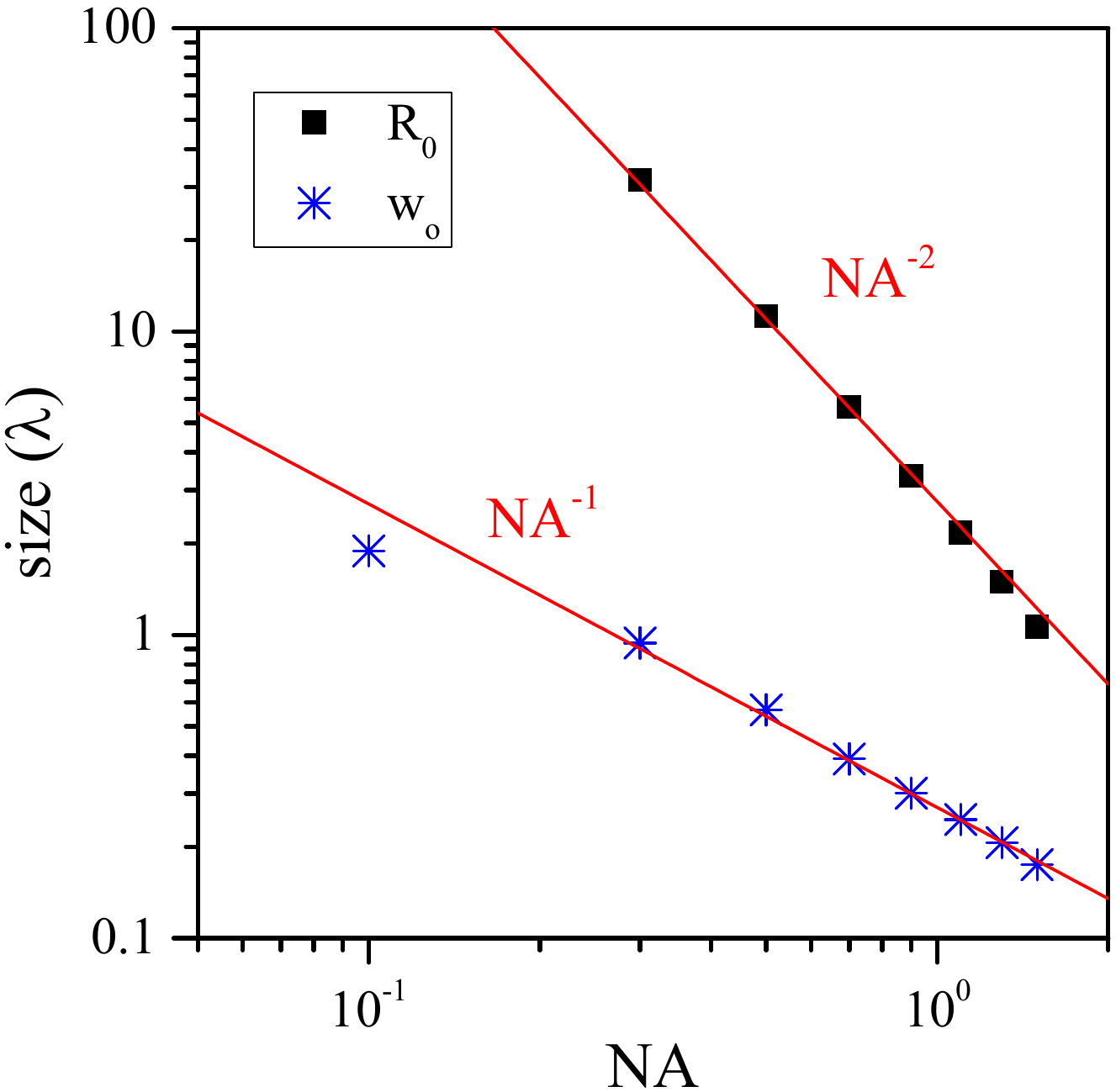}
	\caption{Calculated dependence of the coefficient $R_{0}$ in the
		focal plane (solid squares) on the objective numerical aperture,
		using a filling factor $\beta=0.83$ as in the experiment in a medium of $n=1.5185$. The Gaussian width
		$w_0$ for the in-plane distribution of $\AF^+$ is also shown
		(stars). Sizes are in unit of wavelength ($\lambda$). Red lines show linear and quadratic dependencies on
		$1/{\rm NA}$ as indicated.} \label{Rzero}
\end{figure}

As described in the previous section and in the manuscript, the
four-wave mixing amplitude ratio $\AF^-/\AF^+$ scales quadratically
with the radial coordinates $R$ (for small values of $R$), such that
in first approximation this coordinate can be calculated as
$R=R_{0}\sqrt{\AF^-/\AF^+}$. Fig.\,\ref{Rzero} shows the dependence
of the parameter $R_{0}$ in the focal plane ($z=0$) on the objective
numerical aperture using a
filling factor $\beta=0.83$ as in the experiment, and in a medium of
$n=1.5185$. We find that $R_{0}$ scales quadratically with
$1/{\rm NA}$. The width $w_0$ for the in-plane distribution of
$\AF^+$ fitted as a Gaussian function using the curvature near the
center, i.e. $\AF^+=A_{0}e^{-(R/w_{0})^2}$, is also shown,
representing an effective point-spread function which scales
linearly with $1/{\rm NA}$.

\subsection{Amplitude and phase of FWM ratio versus particle ellipticity}

\begin{figure*}[t!]
	\includegraphics*[width=16cm]{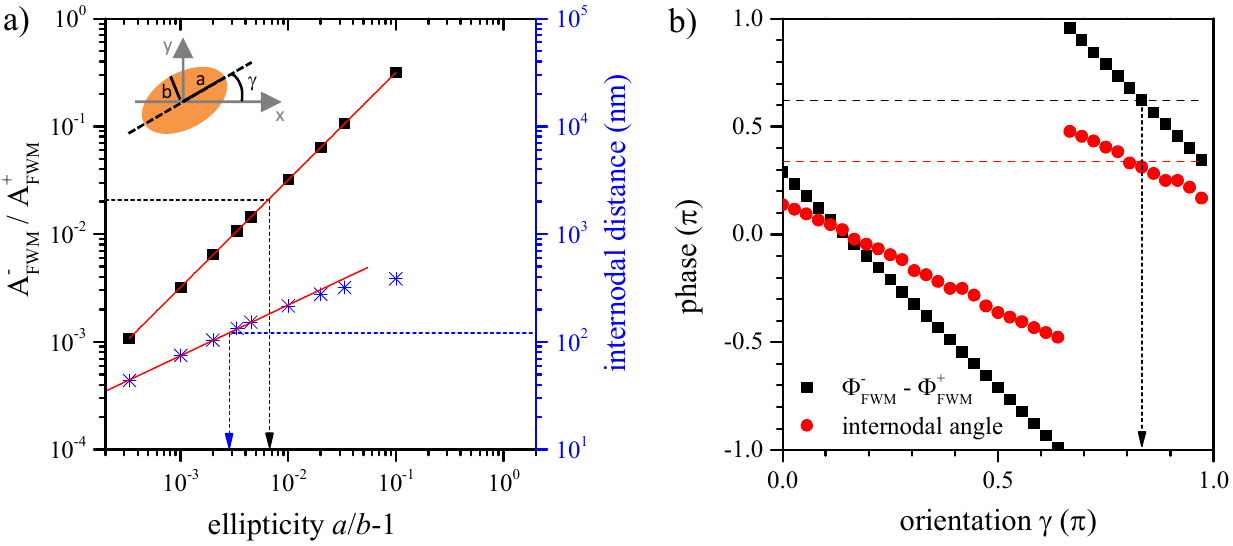}
	\caption{Calculated dependencies of the amplitude and phase of the cross to co-circularly polarised FWM ratio  $\EF^-/\EF^+$  in the focal plane versus
		nanoparticle ellipticity and orientation. a) FWM ratio
		amplitude $\AF^-/\AF^+$ in the focus center (left axis) and distance between the two displaced amplitude minima (right axis) versus particle
		ellipticity $a/b-1$. b) Phase of the FWM ratio $\PhiF^- -
		\PhiF^+$in the focus center (squares) and angle of the internodal axis relative to the $x$-axis (circles). Dashed lines show the experimental values from Fig.\,4 in the main text. The inset in (a) shows a sketch of the
		elliptical nanoparticle with long (short) semi-axis $a$ ($b$), and
		orientation angle $\gamma$. \label{asymmetry}}
\end{figure*}

\begin{figure}[t!]
	\includegraphics*[width=9.0cm]{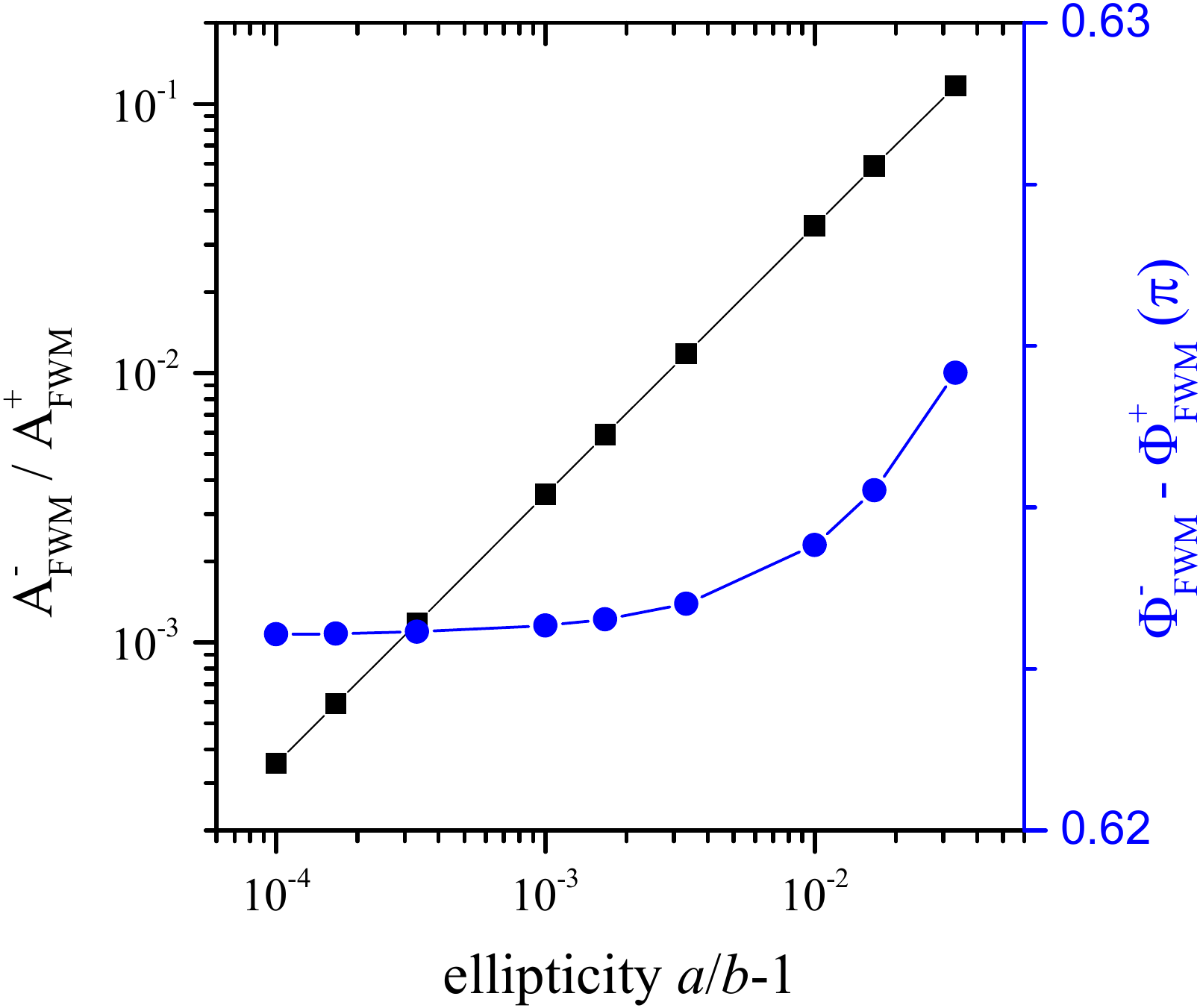}
	\caption{Calculated cross to co-circularly polarised FWM ratio
		amplitude and phase in the focus center versus nanoparticle
		ellipticity, for a particle orientation angle $\gamma=150^\circ$.}
	\label{phase}
\end{figure}

Although being a limitation to achieve nanometric position accuracy,
it is remarkable how sensitive the described FWM technique is to
particle asymmetry, which could be used as a new tool to detect
particle ellipticity down to $10^{-4}$ (corresponding to atomic
accuracy comparable to TEM) as well as particle orientation. This is
shown in \Fig{asymmetry}. The FWM amplitude ratio in the focus center scales
linearly with the particle ellipticity $a/b-1$ (where we assumed
$b=c$ and $a$, $b$ and $c$ aligned along the $x$, $y$ and $z$ axis
respectively). Furthermore the distance between the two displaced
amplitude minima scales with the square root of the ellipticity.
\Fig{asymmetry}a shows these dependencies (red lines) and the
experimental values (dashed lines) observed in Fig.\,4 of the main
manuscript, which indicate an ellipticity of about 0.5\%. The
occurrence of higher order modes in the measured data could also be
used with appropriate modelling (beyond the the Rayleigh regime) to determine more details about the specific
shape of the measured particle. \Fig{asymmetry}b shows that the
finite value of the phase of the FWM ratio in the focus center
is given by -2$\gamma$ plus an offset, with the particle orientation
angle $\gamma$ defined as the angle between the longer particle axis
and the $x$-axis (see sketch in Fig.\,\ref{asymmetry}). Conversely,
this phase is almost independent on the ellipticity (see
\Fig{phase}). Furthermore, the orientation angle of the internodal
axis relative to the $x$-axis scales as -$\gamma$ (plus an offset).
Dashed lines show the experimental values from Fig.\,4 of the main
manuscript which indicate that the particle was oriented with $\gamma$ of about $150^\circ$.

\subsection{Polarisability resulting in an $l=1$ optical vortex of the cross-circularly polarised FWM}

\begin{figure}[t!]
	\includegraphics*[width=16.5cm]{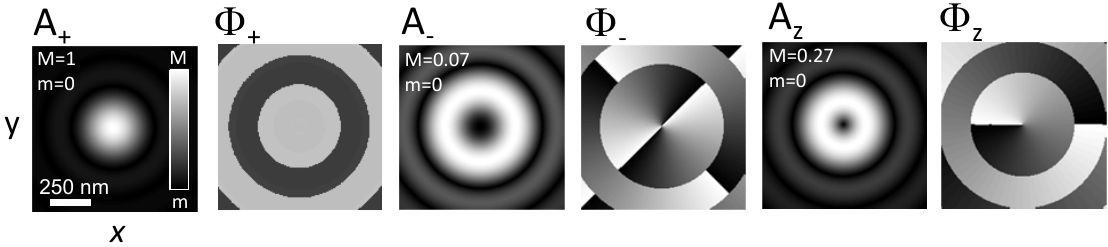}
	\caption{Calculated field distribution in the focal plane of a
		1.45\,NA objective for an incident field left ($\sigma^+$)
		circularly polarized, using a filling
		factor $\beta=0.83$ as in the experiment. $A_+$, $A_-$ and $A_{\rm z}$ are the
		co-circular, cross-circular and longitudinally polarised amplitudes,
		and $\Phi_+$, $\Phi_-$, and $\Phi_{\rm z}$ the corresponding
		phases. Linear grey scale from $-\pi$ to
		$\pi$ for phases, and from $m$ to $M$ for field amplitudes, as
		indicated} \label{figEz}
\end{figure}

Firstly, we note that for a circularly polarized incident field, the
longitudinally polarized field component ($E_{\rm z}$) in the focal
plane exhibits an $l=1$ optical vortex, while the cross-circularly
polarized component is an $l=2$ vortex. This is shown in
Fig.\,\ref{figEz} for the case of a 1.45\,NA objective, as
calculated in the main manuscript in Fig.\,1b.

In order to reproduce the experimental findings in Fig.\,6a of the
main manuscript, we assumed a particle polarisability tensor
representing a rotationally-averaged elliptical particle (hence an
effectively spherical particle with equal semiaxis) plus a
contribution coupling the longitudinal field component into the
$(x,y)$ plane. Therefore, the particle polarisability tensor was
described as:

\be \alphah=\alpha_{0} \left(\begin{matrix} 1&0&0\\
	0&1&0\\0&0&1\end{matrix}\right) +C\alpha_{0}\left(\begin{matrix} 0&0&1\\
	0&0&0\\1&0&0\end{matrix}\right) \ee

with $\alpha_{0}$ given by Eq.\,\ref{eqnalpha} for $a=b=c=25$\,nm in
a surrounding medium consisting of water with index
$n=\sqrt{\epsm}=1.333$, and $C$ being a complex coefficient, the
amplitude of which was adjusted to reproduce the strength of the FWM
amplitude ratio in the experiment (for the calculations in Fig.\,6c
we used $|C|$=0.37).

\section{Experiments}

\subsection{Gold nanoparticles attached onto glass}
\begin{figure}[t!]
	\includegraphics*[width=14.0cm]{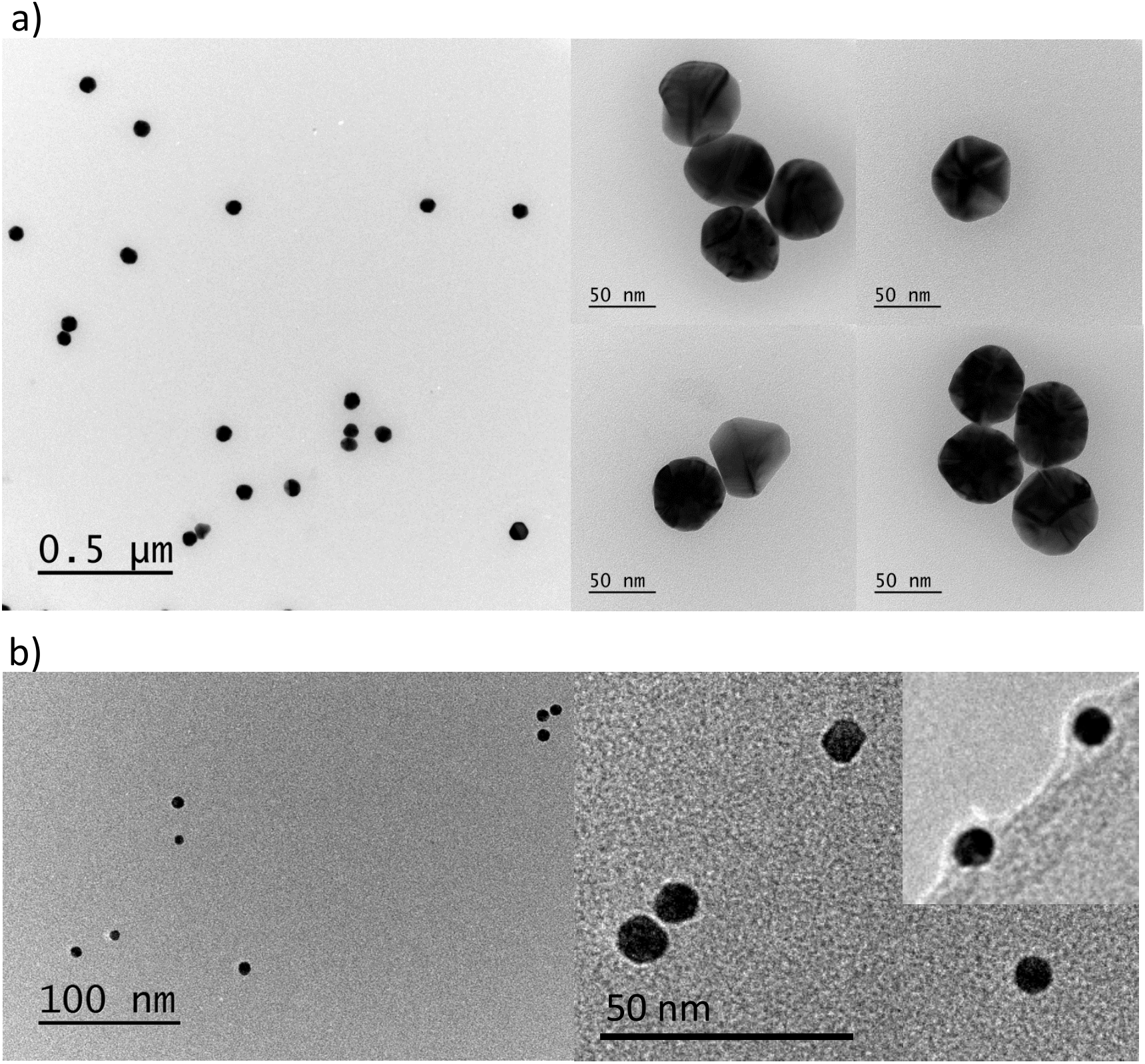}
	\caption{Transmission electron microscopy of nominally spherical
		30\,nm radius (a), and 5\,nm radius (b) gold nanoparticles.}
	\label{TEM}
\end{figure}

Before use, glass slides and coverslips were cleaned from
debris. Cleaning was performed first with acetone and high-quality
cleanroom wipes, followed by a chemical etch called Caro's etch, or
more commonly, Piranha etch. The investigated samples were nominally
spherical gold NPs of 30\,nm and 5\,nm radius (BBI Solutions) drop
cast onto a glass coverslip, covered in silicon oil (refractive
index $n=1.518$) and sealed with a glass slide. Examples of
transmission electron microscopy (TEM) images of these NPs are shown
in Fig.\,\ref{TEM}.

\subsection{Gold nanoparticles in agarose gel}
Before use, glass slides and coverslips were cleaned with acetone.
500\,mg of agar powder (high molecular biology grade agarose,
Bioline Cat. 41025) was placed in a conical glass beaker filled with
10\,mL of pure water, and the mixture was boiled in a microwave for
about 60\,s until a homogeneous liquid was formed. Nominally
spherical gold NPs of 25\,nm radius (BBI Solutions) in pure water
were added to the mixture to achieve a typical concentration of
$10^9$\,NPs/mL. A chamber was formed by attaching an adhesive
imaging spacer of 0.12\,mm thickness with a 13\,mm diameter hole
(GraceBioLabs) onto a glass coverslip. 20\,${\mu}$L volume of the
gold NP agar mixture was pipetted into the chamber and sealed with a
glass slide.

\subsection{Gold nanoparticles in fixed cells}

HeLa cells were seeded onto gridded coverslips and then loaded with
20\,nm radius gold NPs via clathrin mediated endocytosis, using the
transferrin (Tf) ligand attached onto the NP via a
biotin-streptavidin conjugate.  Tf-biotin was purchased from Sigma
Aldrich and 20\,nm radius gold NPs covalently bound to streptavidin
were purchased from Innova Biosciences. Cells were placed on ice for
10\,min to inhibit endocytosis, incubated with 50\,$\mu$g/mL
Tf-biotin in ice-cold serum-free medium for 8\,min and then washed 3
times in ice-cold phosphate-buffered saline (PBS) at pH 7.4. Cells
were then incubated with a dilution of gold NP-streptavidin
conjugate in ice-cold serum-free medium, followed by washing 3 times
in PBS pH 7.4. Cells were subsequently incubated in pre-warmed
imaging medium for 6 hours to enable endocytosis, and then fixed in
3\% paraformaldehyde for 10\,mins at room temperature. They were
washed 3 times in PBS at room temperature and mounted onto a glass
slide using a DAKO (Dako UK Ltd) mounting medium at 80\%.

\subsection{Lock-In detection of the modulation}

A digital lock-in amplifier Zurich Instruments HF2LI was used to
detect the FWM signal at both sidebands $\nu_2\pm\nu_{\rm
	m}-\nu_{\rm L}=(2\pm0.4)$MHz together with the reflected probe at
the carrier $\nu_{\rm c}=\nu_2-\nu_{\rm L}=2$MHz ($\nu_{2}=$82\,MHz
is the frequency up-shift of the probe field via AOM$_2$, $\nu_{\rm
	m}=0.4$\,MHz is the intensity modulation of the pump via AOM$_1$,
and $\nu_{\rm L}=80$\,MHz is the pulsed laser repetition rate, see
also Fig.\,1 in main paper). To produce a modulation of the pump
intensity, we supplied a modulation at $\nu_{\rm m}$ to the digital
input of the AOM$_1$ driver (Intraaction DFE-834C4-6), generated by the internal oscillator of the lock-in. To provide
an electronic reference to the lock-in at the carrier frequency of
2\,MHz, we assembled an external electronic circuit using coaxial
BNC components (Minicircuits), mimicking the optical mixing. The
laser repetition frequency was taken from the 80\,MHz photodiode
signal output of the Ti:Sa laser (Newport/Spectra Physics MaiTai).
After filtering for the first harmonics using a 50\,MHz high pass
(BHP-50+) and a 100\,MHz low-pass (BLP-100+), the signal was mixed
with the frequency $\nu_2$ from the +10dBm reference output of the
AOM$_2$ driver (Intraaction DFE-774C4-6) using an electronic mixer
(ZP-1LH+) and a 10\,MHz low pass filter (BLP-10.7+) to produce an
electric sine wave at the difference frequency $\nu_2-\nu_{\rm
	L}=2$\,MHz. The latter was then converted from sine to TTL using a
comparator (Pulse Research Lab - PRL-350TTL-220), to be suited as
external reference for the HF2LI using the DIO0 input (the two
available high-frequency analog inputs were used for the two
balanced-diode signals).

%\begin{figure}
%\includegraphics*[width=10.0cm]{FigSM1}
%\caption{Electronic circuit diagram for modulation and lock-in
%detection of heterodyne FWM.} \label{fig:electronics}
%\end{figure}

To discuss the analysis of the dual sideband detection by the
lock-in, we start by describing the detected voltage of the
balanced-diode signal as the real part of
\be U=a_{\rm c} e^{i\omega_{\rm c} t}\left( 1+\am \cos(\omm t +
\varphi)\right). \ee
with the complex carrier amplitude $a_{\rm c}$, which is measured
directly by the ZI-HF2LI via the in-phase ($Re$) and in-quadrature ($Im$) components at the
carrier frequency $\nu_{\rm c}=\omega_{\rm c}/(2\pi)$, and the complex
relative modulation of the carrier $\am$ with the frequency
$0<2\pi\nu_{\rm m}=\omm<\omega_{\rm c}$ and the phase $\varphi$. The
modulation is due to the intensity modulation of the pump field, which is a real quantity and can thus be described by a cosine. This can be rewritten as
\be U=a_{\rm c} e^{i\omega_{\rm c} t}+\frac{a_{\rm
		c}\am}{2}e^{i\varphi} e^{i(\omega_{\rm c}+\omm) t}+\frac{a_{\rm
		c}\am}{2}e^{-i\varphi} e^{i(\omega_{\rm c}-\omm)t}. \ee
Accordingly, the complex amplitudes of the lower and upper sidebands
are $ a_\pm =a_{\rm c}\am e^{\pm i\varphi}/2 $, which are detected
together with $a_{\rm c}$ by the ZI-HF2LI using the dual sideband
modulation detection. The modulation phase $\varphi$ can be
determined modulo $\pi$ by
\be  e^{2i\varphi} =  \frac{a_+}{a_-}\ee
and is due to the delay $\tau$ between the modulation output of the
lock-in and the detection of the amplitude modulation by the
lock-in, specifically $\varphi=-\omm\tau$, resulting in the
modulation term $\cos(\omm(t-\tau))$. The modulation of the detected
probe field, which is the FWM signal, is given by
\be  a_{\rm c}\am =  a_+ e^{-i\varphi}+a_- e^{i\varphi}
\label{eqn:a0am}\ee

The determination of $\varphi$ modulo $\pi$ leads to an uncertainty
of the sign of the modulation. The phase shift $\varphi$ can be
determined including its sign by directly detecting the reflected
pump beam which is frequency up-shifted via AOM$_1$ by the amount
$\nu_1=83$\,MHz, using $\nu_1-\nu_{\rm L}$ instead of
$\nu_2-\nu_{\rm L}$ in the lock-in reference, and knowing that
$\am>0$. Since the phase shift is given by electronic and
acousto-optic delays it is stable to a few degrees. Once $\varphi$
is known, the four-wave mixing field $\am a_{\rm c}$ is determined
using \Eq{eqn:a0am}.

\subsection{Shot-noise limited detection: Dependence on reference power}

\begin{figure}[t!]
	\includegraphics*[width=7cm]{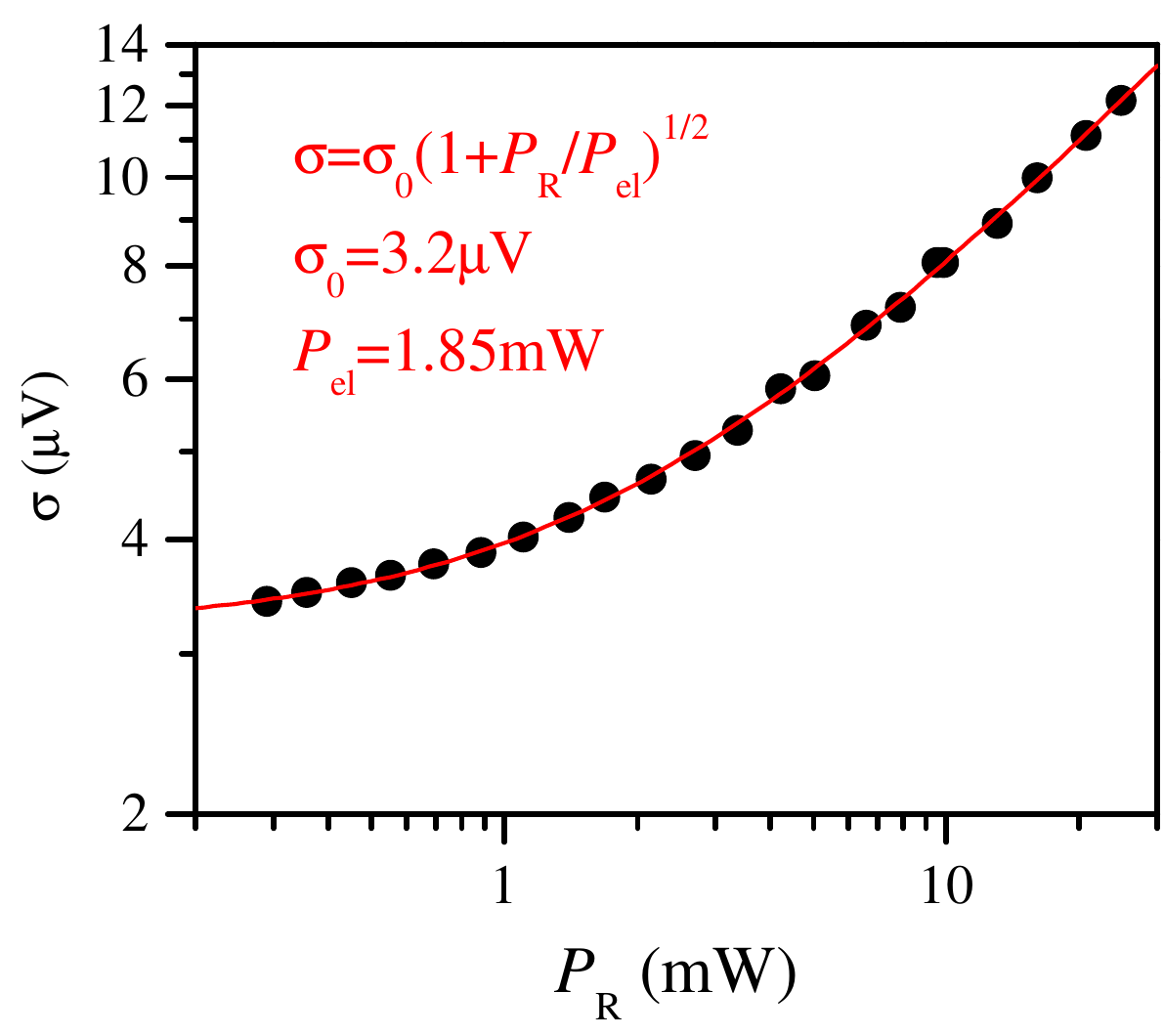}
	\caption{Standard deviation $\sigma$ of the distribution of the
		measured FWM field (in-phase $Re$ component, cross-circularly polarized) in a
		spatial region away from the particle, as a function of the total
		power in the reference beam. 1\,ms pixel dwell time. A fit to the data is shown as indicated}
	\label{refpower}
\end{figure}

The experimental noise was evaluated by taking the statistical
distribution of the measured FWM field (in both $Re$ and $Im$
components) in a spatial region away from the particle where no
detectable signal is present. The standard deviation $\sigma$ of
this distribution was deduced for each component. For the measurement conditions as in Fig.\,4 in the main paper, $\sigma$ was found to be the identical in the $Re$ and $Im$ component, and for the co-polarised and
cross-polarised components, as expected for an experimental noise
given by the shot-noise of the reference beam and the electronic noise. \Fig{refpower} shows $\sigma$  as a function of the total power $\Pref$ of the reference beam entering the balanced detector, which is split onto the 4 photodiodes (Hamamatsu S5973-02) of the dual-polarization balanced detector with a transimpedance of $10^4$\,V/A. It shows a noise at a frequency of  2\,MHz and 1\,ms integration time of $\sigma_0\sqrt{1+\Pref/\Pel}$, with  the electronic noise of $\sigma_0=3.2\,\mu$eV, equivalent to the shot noise created by $\Pel=1.85$\,mW at the detector input, and is thus shot-noise limited for $\Pref>\Pel$. We used $\Pref\sim3\,$mW in the experiments shown.

\subsection{Pump-induced change of particle polarisability: Pump
	power and delay dependence}

\begin{figure}[t]
	\includegraphics*[width=14cm]{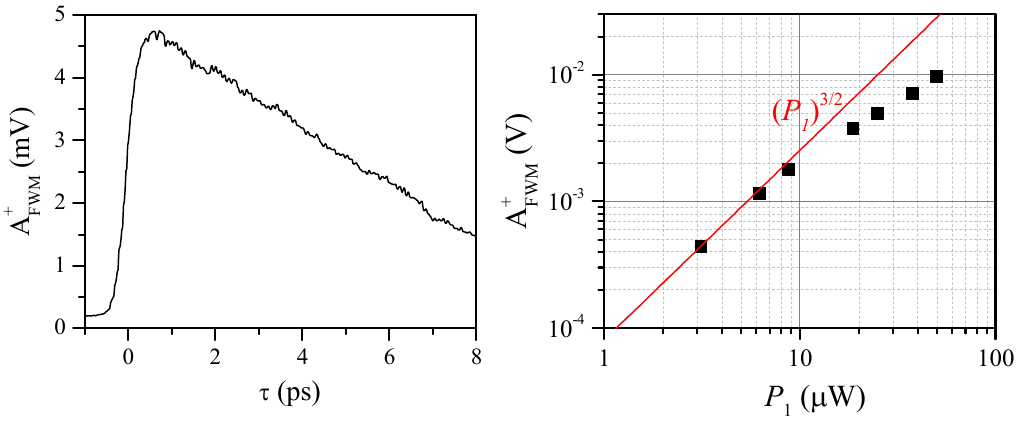}
	\caption{Left: Co-polarised FWM field amplitude for a 30\,nm radius
		NP at the focus center versus pump-probe delay time $\tau$ for a
		pump (probe) power at the sample of 40\,${\mu}$W (20\,${\mu}$W).
		Right: Co-polarised FWM field amplitude at the focus center versus
		versus pump power at $\tau=0.5$\,ps. The probe power is half the pump power. The  scaling $\propto P_1^{3/2}$ expected in
		the third-order regime is shown.} \label{pumppower}
\end{figure}

An example of the co-polarised FWM field amplitude $\AF^+$ for a
30\,nm radius NP at the focus center as a function of the pump-probe
delay time and power is shown in Fig.\ref{pumppower}.

As discussed in detail in our previous work \cite{MasiaPRB12}, the
pump-probe delay dependence reflects the transient change of the
electron and lattice temperature following the absorption of the
pump pulse by the nanoparticle. As a result of the ultrafast heating
of the electron gas, the pump-induced change of the particle
polarisability, and in turn the transient FWM field amplitude,
reaches its maximum at a pump-probe delay $\tau$ of about 0.5\,ps.
The subsequent decay on a picosecond time scale reflects the
thermalization with the lattice, i.e. the cooling of the electron
gas via electron-phonon coupling.

The scaling of $\AF^+$ at $\tau=0.5$\,ps with increasing pump and
probe power shown in \Fig{pumppower} follows the expected
third-order nonlinear behavior for small powers, and the onset of
saturation at higher power. Pump power ($P_1$) and probe power
($P_2$) were increased while keeping a constant ratio $2P_{2} = P_1$. At small powers, in the limit of a pump-induced
change of the particle polarisability linear with the pump
intensity, the FWM field amplitude scales as $P_{1} \sqrt{P_2}
\propto P_{1}^{3/2}$.

\subsection{Background-free FWM detection in cells}

To exemplify that our FWM detection is very specific to gold NPs and
free from background even in highly scattering and fluorescing
environments, Fig.\,3 in the main manuscript shows HeLa cells that
have internalized 20\,nm radius gold NPs bound to the transferrin
receptor, via clathrin mediated endocytosis. Cells were fixed (with
3\% paraformaldehyde solution) onto gridded glass coverslips, and
immersed in a mounting medium (80\% Dako) chambered with a glass
slide. They were imaged with FWM using the $100\times$ magnification
1.45\,NA oil-immersion objective specified in the main manuscript.
Differential interference contrast (DIC) microscopy was also
available in the same instrument, with wide-field illumination via
an oil condenser of 1.4\,NA and detection with a sCMOS camera
(Pco.edge 5.5).

\begin{figure}
	\includegraphics*[width=16cm]{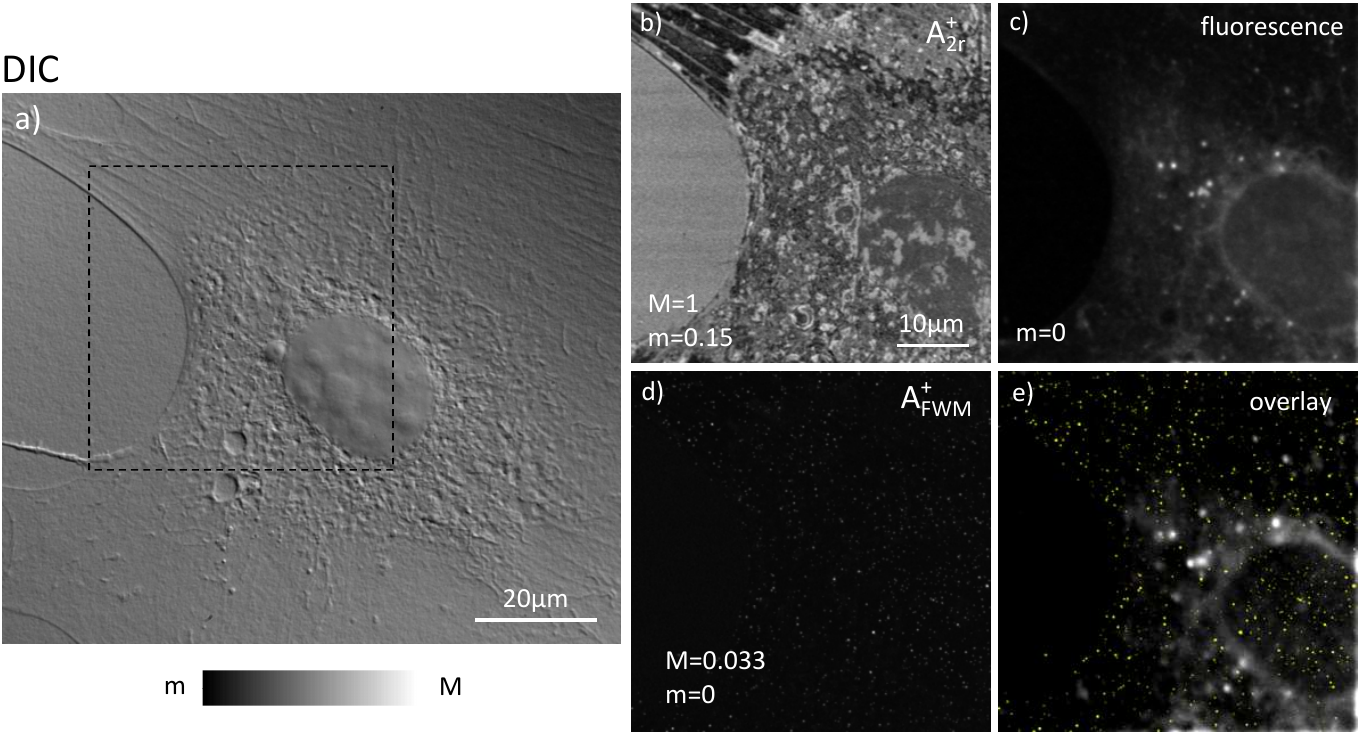}
	\caption{Fixed 3T3 cells that have internalized gold NPs of 10\,nm
		radius imaged by (a) differential interference contrast (DIC)
		microscopy, (b) co-circularly polarised reflection amplitude
		$\Ar^+$, (c) confocal epi-fluorescence, (d) co-circularly polarised
		FWM amplitude $\AF^+$. Amplitude scales from m to M are indicated
		(for confocal fluorescence M$=1.37\times10^6$ (photoelectrons/sec)).
		FWM was acquired with pump-probe delay time of 0.5\,ps, pump (probe)
		power at the sample of 60\,$\mu$W (30\,$\mu$W), 0.4\,ms pixel dwell
		time, pixel size in plane of 63\,nm and z-stacks over 3\,$\mu$m in
		500\,nm z-steps. FWM is shown as a maximum intensity projection over
		the z-stack, while the reflection and fluorescence images are on a
		single $(x,y)$ plane (scanning the sample position). e) is a  false
		color overlay of the FWM (yellow) and fluorescence images in c) and
		d) respectively, brightness and contrast enhanced for visualization
		purposes.} \label{correlative}
\end{figure}

Notably, FWM acquisition can be performed simultaneously with
confocal fluorescence microscopy for correlative co-localisation
analysis. An example of this simultaneous acquisition is shown in
\Fig{correlative}. Confocal epi-fluorescence detection was
implemented in the same microscope set-up, by conjugating the sample
plane onto a confocal pin-hole of adjustable opening in front of a
photomultiplier detector (Hamamatsu H10770A-40). Excitation occurred
via the same laser beam used for FWM; fluorescence was collected via
the same objective (in this case 1.45\,NA oil-immersion) and
spectrally separated using a dichroic beam splitter for pick-up and
a bandpass filter transmitting in the 600-700\,nm wavelength range
in front of the photomultiplier. \Fig{correlative} shows 3T3 cells
loaded with 10\,nm radius gold NPs using a protocol similar to the
one described above for HeLa cells. Fixed cells were image with DIC,
reflection, FWM and simultaneous confocal fluorescence. The
co-circularly polarized reflection image $\Ar^+$ in
\Fig{correlative}b correlates with the cell contour seen in DIC in
the region highlighted by the dashed frame, and shows a spatially
varying contrast due to thickness and refractive index
inhomogeneities in the sample. Epi-fluorescence in
\Fig{correlative}c correlates with the cell morphology seen in DIC
and reflection image, and is dominated by autofluorescence near the
cell nucleus. The co-circularly polarised FWM amplitude $\AF^+$
shown in \Fig{correlative}d is a maximum intensity projection over a
3\,${\mu}$m $z$-stack and clearly indicates the location of single
gold NPs in the cell. FWM imaging is background-free (throughout the
$z$-stack) despite the significant scattering and autofluorescing
cellular contrast seen in \Fig{correlative}a,b,c. \Fig{correlative}e
shows a false color overlay of the FWM and fluorescence image.

\subsection{FWM field images of various particles}

\begin{figure}
	\includegraphics*[width=16.5cm]{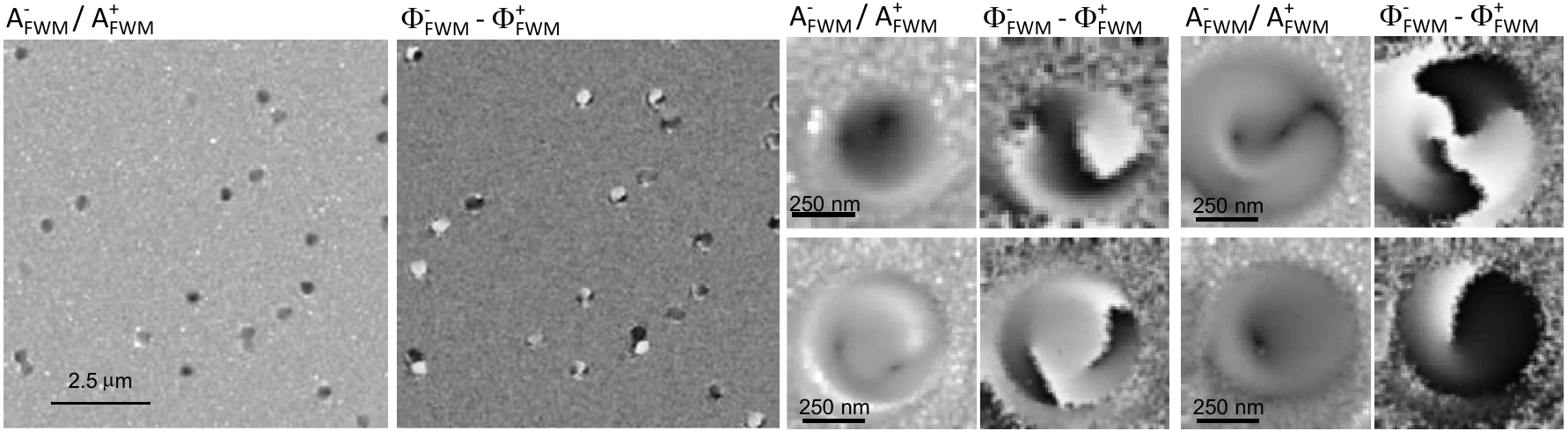}
	\caption{Measured FWM ratio amplitude and phase as a function of NP position in the
		focal plane on various nominally spherical 30\,nm radius single gold
		nanoparticles. Grey scale is from $-\pi$ to $\pi$ for all phases.
		Amplitude ratio is in logarithmic scale over 3 orders of magnitudes
		from 0.01 to 10. Reference power was 2.6\,mW. Pump-probe delay time
		was 0.5\,ps. Pump (probe) power at the sample was 24\,$\mu$W
		(12\,$\mu$W) for (a)-(c), 8\,$\mu$W (4\,$\mu$W) for (d), and
		18\,$\mu$W (9\,$\mu$W) for (e). Pixel dwell time was 0.5\,ms for (a)
		and 3\,ms for (b)-(e). Pixel size in plane was 32\,nm for (a), 17
		for (b) and 10\,nm for (c)-(e).} \label{otherNPs}
\end{figure}

\Fig{otherNPs} gives examples of the measured FWM field ratio $\EF^-
/ \EF^+$ in amplitude and phase in the focal plane for different NPs
of nominally spherical 30\,nm radius, showing a variety of patterns
in terms of the amplitude and phase in the focus center, the
position of the amplitudes nodes, their distance and orientation and
the angular distribution of the ratio phase. \Fig{otherNPs}a is an
overview while \Fig{otherNPs}b-e are higher resolution zooms over
individual NPs. The heterogeneity of the observed patterns is
consistent with the heterogeneous NP shape and asymmetry observed in
TEM.

\subsection{FWM measurements on 5\,nm radius nanoparticles}

\begin{figure}[t!]
	\includegraphics*[width=8cm]{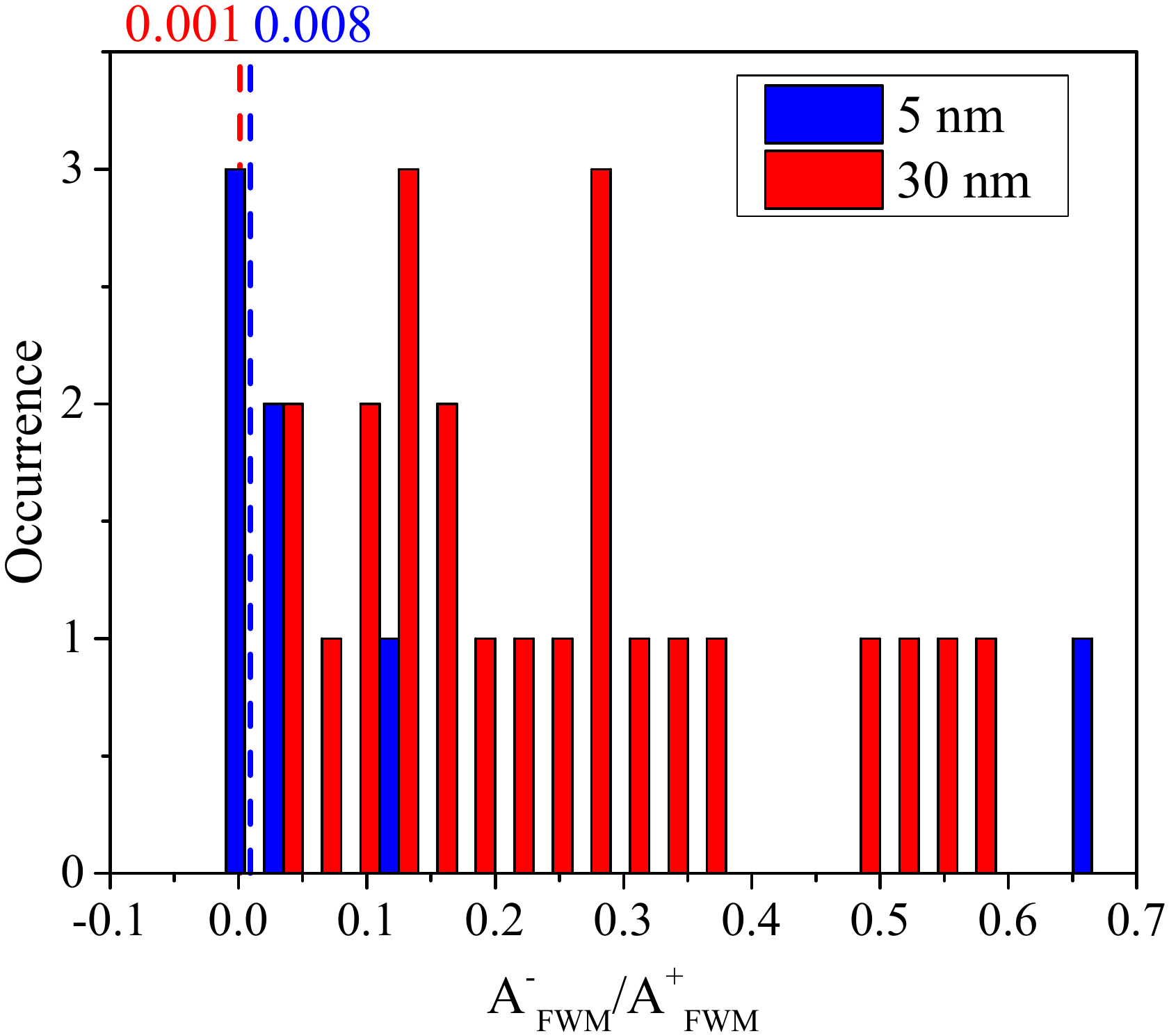}
	\caption{Histogram of the occurrence of the FWM amplitude ratio
		$\AF^-/\AF^+$  measured on
		nominally spherical 5\,nm and 30\,nm radius gold NPs in the focus center, as indicated.
		Dashed lines show the shot-noise limit in the corresponding
		experiments. For the 5\,nm radius NPs, measurements were performed
		typically with 90\,${\mu}$W (45\,${\mu}$W) pump (probe) power at the
		sample, 10\,ms pixel dwell time, 10\,nm pixel size. FWM ratios (and
		corresponding shot-noise) were calculated by spatial averaging over
		an effective area of $9 \times 9$ pixels. For the 30\,nm radius NPs,
		measurements were performed with 24\,${\mu}$W (12\,${\mu}$W) pump
		(probe) power at the sample, 0.5\,ms pixel dwell time, 32\,nm pixel
		size. FWM ratios (and corresponding shot-noise) were calculated by
		spatial averaging over an effective area of $3 \times 3$ pixels.}
	\label{FWM10nm}
\end{figure}

The FWM amplitude ratio $\AF^-/\AF^+$ in the focus center was
measured on nominally spherical 5\,nm radius gold NPs of the type
shown in \Fig{TEM}b, and compared with the ratio obtained on
nominally spherical 30\,nm radius gold NPs of the type shown in
\Fig{TEM}a. The results are summarized in \Fig{FWM10nm}. The
shot-noise limit $\sigma/A_0$ in these experiment is indicated by
the dashed lines. Notably, we observe that for the 5\,nm radius NPs,
a large proportion ($\sim70\%$) has a FWM amplitude ratio
$\AF^-/\AF^+ \leq0.02$ in the center of the focal plane, in this
case limited by the signal-to-noise ratio rather than the particle
asymmetry. Conversely, almost all NPs of nominal 30\,nm radius have
$\AF^-/\AF^+ > 0.02$ which is well above the shot-noise limit and is
due to the particle asymmetry.

\subsection{Parameters for position localisation of a freely rotating single NP}

\begin{figure*}[t]
	\includegraphics*[width=16.5cm]{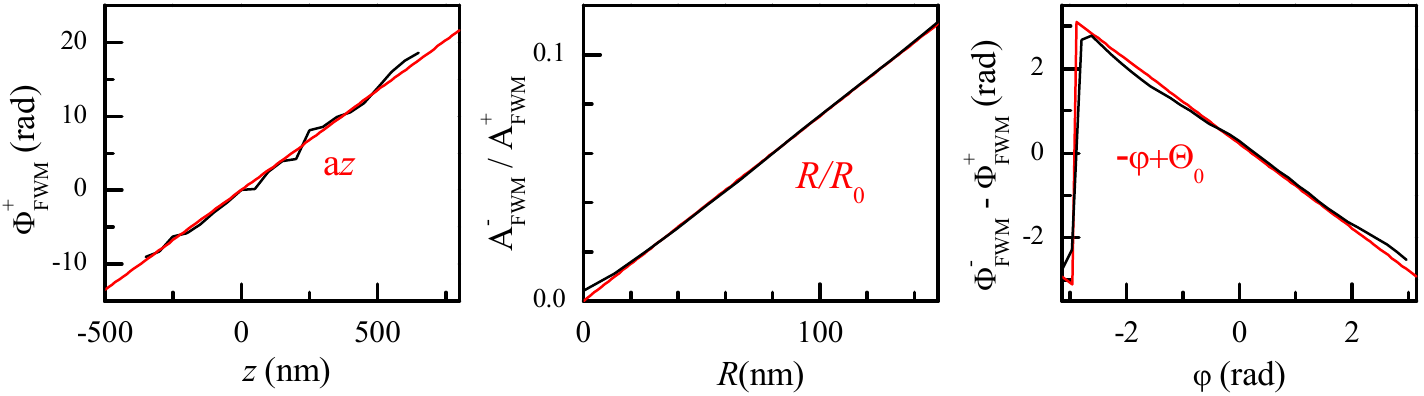}
	\caption{Left: Axial dependence of the phase of the co-polarised FWM
		component (black) together with a linear fit (red line) having the
		slope $\partial z/\partial\Phi=37$\,nm/rad. Middle: Radial
		dependence of the cross to co-polarised FWM amplitude ratio at the
		focal plane $z=0$. The red solid line shows a fit assuming the
		linear dependence $R/R_0=\AF^-/\AF^+$ with $R_0=1.33\,\mu$m at the
		focal plane. Right: Phase of the cross to co-polarised FWM ratio
		$\Theta=\PhiF^- - \PhiF^+$ versus polar angle in the $z=0$ plane,
		following the relationship $-\varphi=\Theta-\Theta_0$. The red solid
		line shows a fit using $\Theta_0=0.22$. \label{figlocch}}
\end{figure*}

From the measurements on a freely rotating NP in an agarose gel
shown in the main manuscript in Fig.\,6 we have deduced the
relationship between the FWM amplitude ratio $\AF^-/\AF^+$ and phase
$\PhiF^- - \PhiF^+$ and the position coordinates in the focal plane,
as measured from the sensor positions in the scanned piezoelectric
sample stage. This is shown in Fig.\,\ref{figlocch}. The $(x,y)$
experimental scans in Fig.\,6 were first converted in 2D images
along the polar coordinates $R$, $\varphi$. The data for
$\AF^-/\AF^+$ were averaged over the full range of the $\varphi$
coordinate to obtain the one-dimensional dependence versus $R$ shown
in Fig.\,\ref{figlocch}. Similarly, the data for $\PhiF^- - \PhiF^+$
were averaged in $R$ over a region of about $60$\,nm\,$<R<100$\,nm
to obtain the dependence versus $\varphi$. The FWM field phase
$\PhiF^+$ versus $z$ also shown in Fig.\,\ref{figlocch} was taken
from rapid 3D scans. We find that $\PhiF^+$ varies linearly with
$z$, as discussed in Section S1.iv, with the slope $\partial
z/\partial\Phi=37$\,nm/rad (note that the experiments in the main
manuscript in Fig.\,6 were performed using a 1.27\,NA
water-immersion objective). We also find that $\AF^-/\AF^+$ has a
simple linear dependence in $R$ given by $R/R_0=\AF^-/\AF^+$ with
$R_0=1.33\,\mu$m in the focal plane. Moreover, the phase of the
cross to co-polarised FWM ratio $\Theta=\PhiF^- - \PhiF^+$ versus
polar angle follows the relationship $-\varphi=\Theta-\Theta_0$ with
$\Theta_0=0.22$ for the data shown in Fig.\,\ref{figlocch}.

\subsection{Axial fluctuations of a rotating single NP encaged in an agarose gel pocket}

\begin{figure*}
	\includegraphics*[width=12cm]{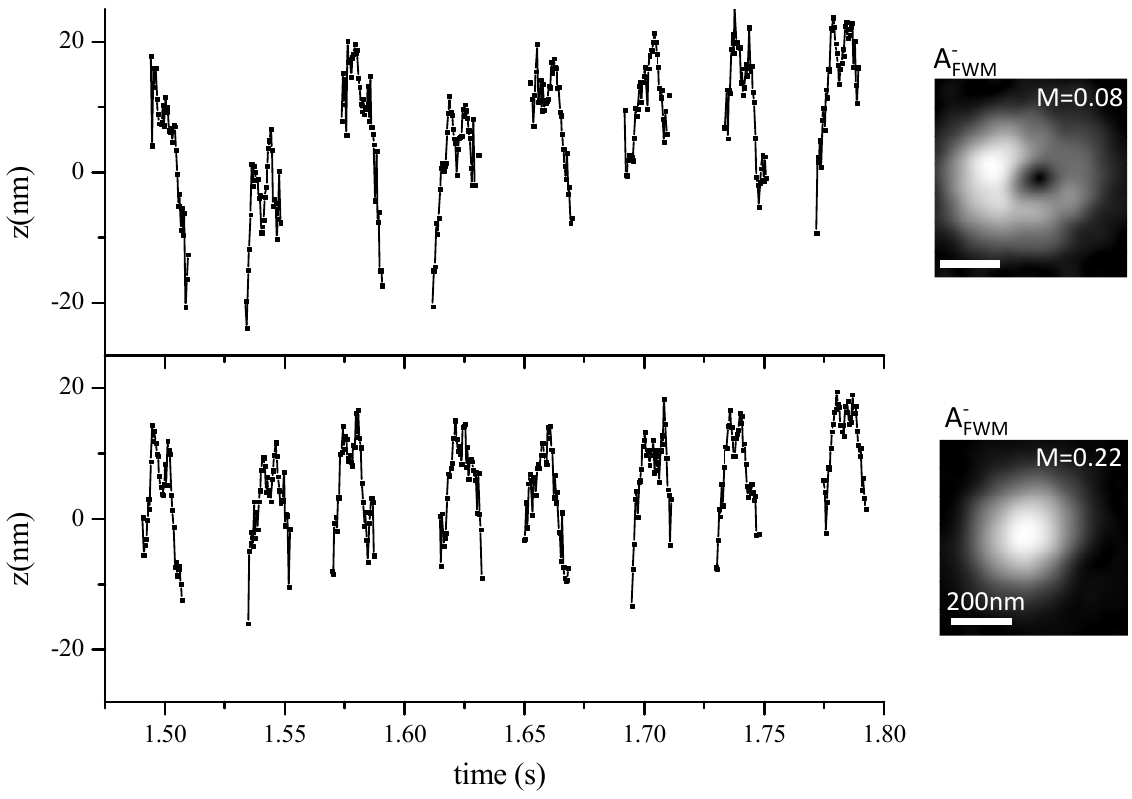}
	\caption{Time traces of the NP axial position coordinate retrieved
		from the measured $\PhiF^+$ during the scan corresponding to the
		distribution of the cross-circularly polarized FWM amplitude $\AF^-$
		shown in the inset. Images are on a linear grey scale from 0 to $M$,
		with $M$ given relative to the maximum $\AF^+$. Pump (probe) power
		at the sample was 70\,${\mu}$W (10\,${\mu}$W); the focused size of
		the pump beam is twice larger than the probe in this experiment.
		Measurements were performed with 0.5\,ms pixel dwell time, 0.5\,ps
		pump-probe delay time, and 13\,nm pixel size
		in-plane.\label{Zfluct}}
\end{figure*}

We investigated the fluctuations in the axial position coordinate
retrieved from the phase $\PhiF^+$ of the co-circularly polarised
FWM during a scan giving rise to an $l=1$ optical vortex in the
cross-circularly polarized FWM (as shown in the main manuscript in
Fig.\,6b) in comparison with those observed during a scan where the
highest value of $\AF^-$ in the focus is measured. This is shown in
\Fig{Zfluct}. The behaviour is quite similar on both traces,
dominated by systematic effects in the $z$ coordinate due to a
cross-coupling to the $xy$ scanning, and a slow drift in $z$ -- note
that the $z$ range shown is only 50\,nm. Superimposed, we find
random variations in $z$ which are larger than the shot noise
(estimated to be $\leq3$\,nm under the experimental conditions in
\Fig{Zfluct}). Based on our calculation of the effect of NP
asymmetry on the cross-circularly polarized FWM (see also section
S1.vi), the trace corresponding to the large $\AF^-$ is attributed
to the NP not rotating but stuck in a position where the NP
ellipticity determines $\AF^-$ in the focus center (from which we
estimate the NP ellipticity to be $a/b-1=0.08$). For the trace
corresponding to the observation of the $l=1$ mode in $\AF^-$, we do
observe larger jumps in the $z$-coordinate and the range spanned is
slightly larger, consistent with the idea that the NP is freely
rotating in 3D in this case, and hence also bouncing about within
the agarose gel pocket in which it is trapped.

\begin{figure*}
	\includegraphics*[width=10cm]{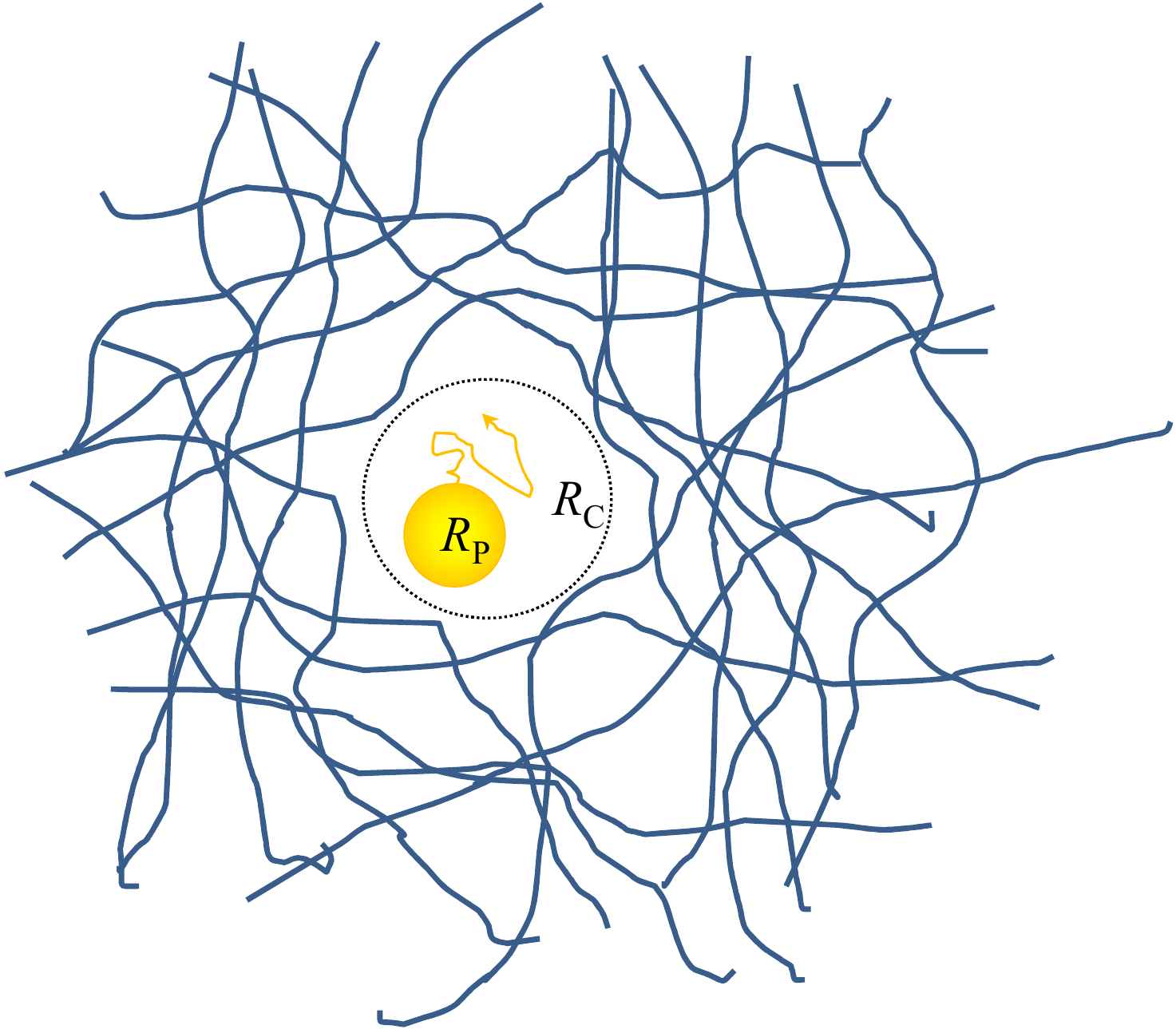}
	\caption{Sketch of the diffusion of a NP caged in a pocket of agar. The NP radius $\RNP$ and the cage radius $\RC$ is indicated, together with a possible diffusion path.\label{Diffusion}}
\end{figure*}

\begin{figure*}
	\includegraphics*[width=10cm]{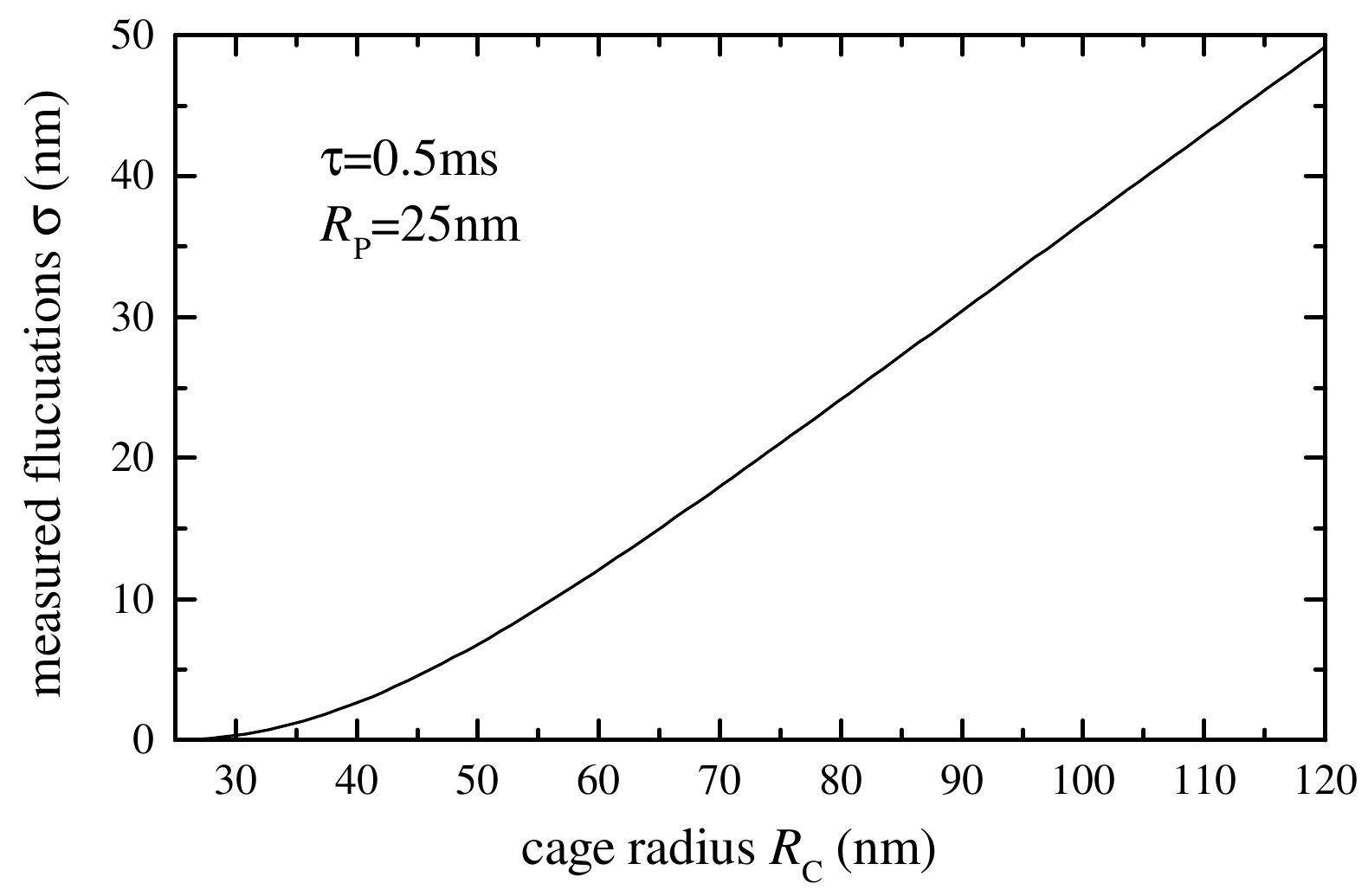}
	\caption{Estimated standard deviation of the measured position of a NP of
		radius $\RNP=25\,$nm diffusing in a cage of radius $\RC$ in water at
		room temperature for an integration time of
		$\tau=0.5\,$ms.\label{DiffusionS}}
\end{figure*}

To estimate the effect of restricted diffusion of a NP confined in a
cage in agar, as sketched in \Fig{Diffusion}, we assume a NP radius
of $\RNP$ and a cage radius of $\RC$. The free diffusion of the NP
in water is given by
\be \langle x^2 \rangle = 2Dt = \frac{\kB T}{3\pi\eta\RNP}t\ee
for each dimension of space, where $\kB$ is Boltzmann's constant,
$T$ is the temperature, and $\eta$ is the dynamic viscosity of
water. Here we are interested only in the $z$ coordinate. The
boundary of the cage will limit this diffusion, confining the
particle motion to a range $2(\RC-\RNP)$, which the particle
explores in the characteristic time
\be \tauC=\frac{12\pi\eta\RNP(\RC-\RNP)^2}{\kB T} \ee
If the measurement integration time $\tau$ is smaller than $\tauC$,
the measured position fluctuation will explore the full cage size.
If instead $\tau>\tauC$, the measured position will be an averaged
position, with less fluctuations, an effect also known as motional
narrowing. The standard deviation of the position fluctuations
scales as the inverse root of the number of averages, hence is given
by
\be \sigma=\frac{2(\RC-\RNP)}{\sqrt{12(1+\tau/\tauC)}} \ee
where the $1/\sqrt{12}$ factor originates from the standard
deviation of a uniform distribution. For the present measurements,
we have $T=300\,$K, $\eta=1$\,mPa$\cdot$s, $\RNP=25\,$nm,
$\tau=0.5\,$ms. The resulting $\sigma$ as function of the cage
radius is given in \Fig{DiffusionS}. We find that due to the
motional narrowing, $\sigma$ is suppressed to below 5\,nm for $\RC
\lesssim 45\,$nm, for which the NP has 40\,nm space to move and thus
also to rotate. The observed fluctuations in \Fig{Zfluct}, which
have a $\sigma$ in in the order of 5\,nm, are therefore consistent
with the assumption of free rotational diffusion.

\subsection{Randomly oriented non-rotating particle}

\begin{figure*}[t!]
	\includegraphics*[width=16.5cm]{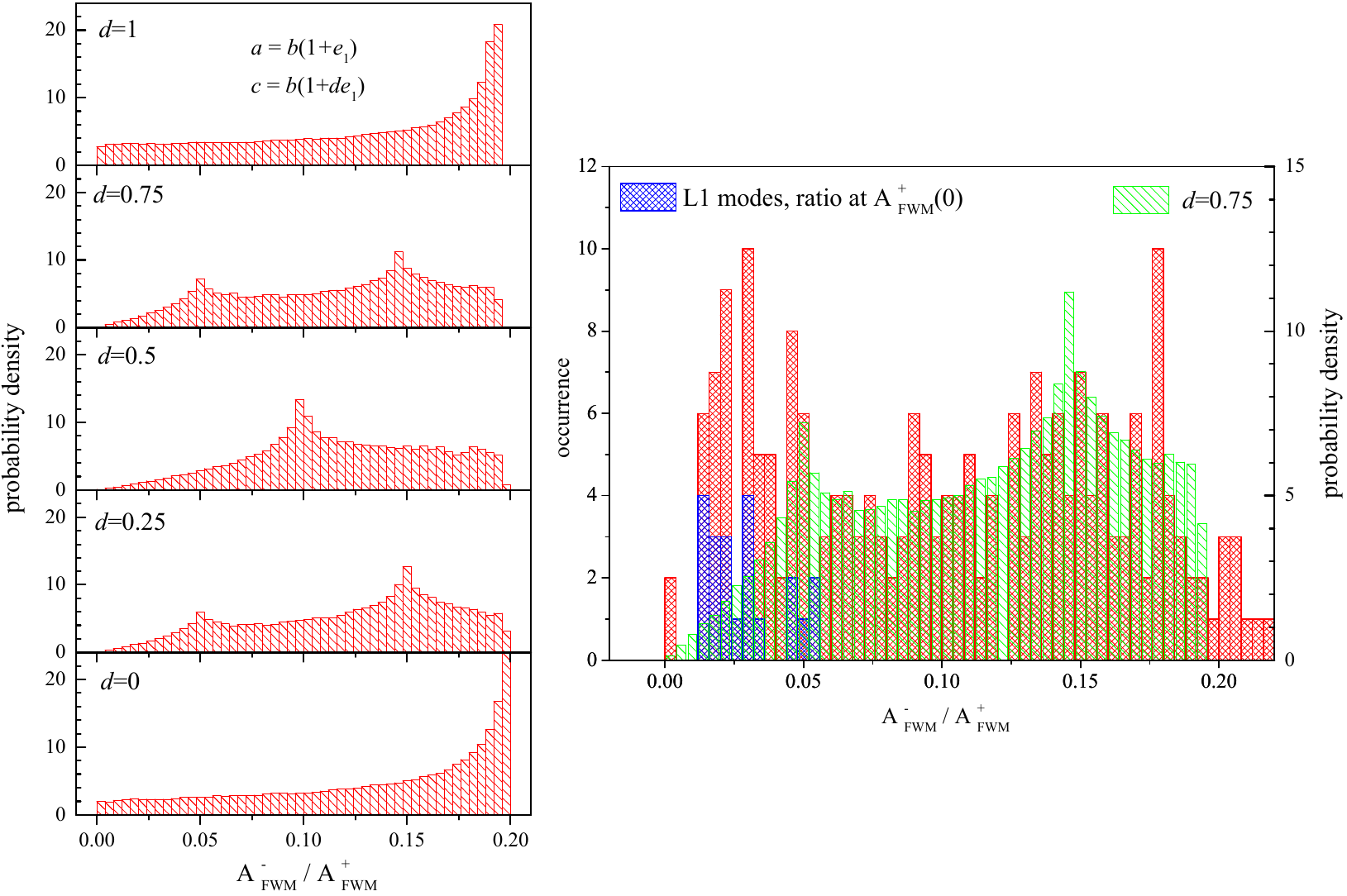}
	\caption{Left: Calculated histograms of the FWM ratio amplitude in
		the focus center for an asymmetric ellipsoid stuck in randomly
		oriented positions. The ellipsoid has semiaxis $a=b(1+e_1)$ and
		$c=b(1+de_1)$, with $e_1=0.0825$, and $d$ is varied as indicated.
		Right: Histogram of the FWM ratio amplitude in the focus center
		experimentally measured while repeatedly scanning over time a single
		25\,nm radius NP in agarose gel in the focal plane. The blue bars
		represent scans where an $l=1$ optical vortex was observed in the
		measurement of $\AF^-$ with its node slightly off-centered, hence
		the finite non-zero value of $\AF^-/\AF^+$. Green bars are the
		calculated histogram shown in left for $d=0.75$.\label{asymhisto}}
\end{figure*}

We cannot reproduce the experimental finding of an $l=1$ optical
vortex in the cross-circularly polarized FWM by assuming a
non-rotating randomly-oriented asymmetric particle. This is shown in
Fig.\,\ref{asymhisto} where we have calculated the statistical
distribution of the FWM ratio $\AF^-/\AF^+$ in the focus center
assuming an asymmetric particle ellipsoid with axis $a=b(1+e_1)$ and
$c=b(1+de_1)$ using the ellipticity $e_1=0.0825$ and considering
random orientations in 3D. Note that by varying the parameter $d$
from 1 to 0 we change the particle shape from oblate to prolate. The
histograms shown in Fig.\,\ref{asymhisto} are significantly
different from the experimentally observed one, the latter
exhibiting a large occurrence of FWM ratios having values
$\AF^-/\AF^+<0.05$ in the range where we see $l=1$ optical vortices.
These calculations indicate that the experimental findings of an
$l=1$ optical vortex in $\AF^-$ cannot be attributed to a
non-rotating asymmetric particle stuck in a randomly-oriented
position.

\subsection{{\it In situ} calibration of in-plane particle position}

\begin{figure*}[t!]
	\includegraphics*[width=8cm]{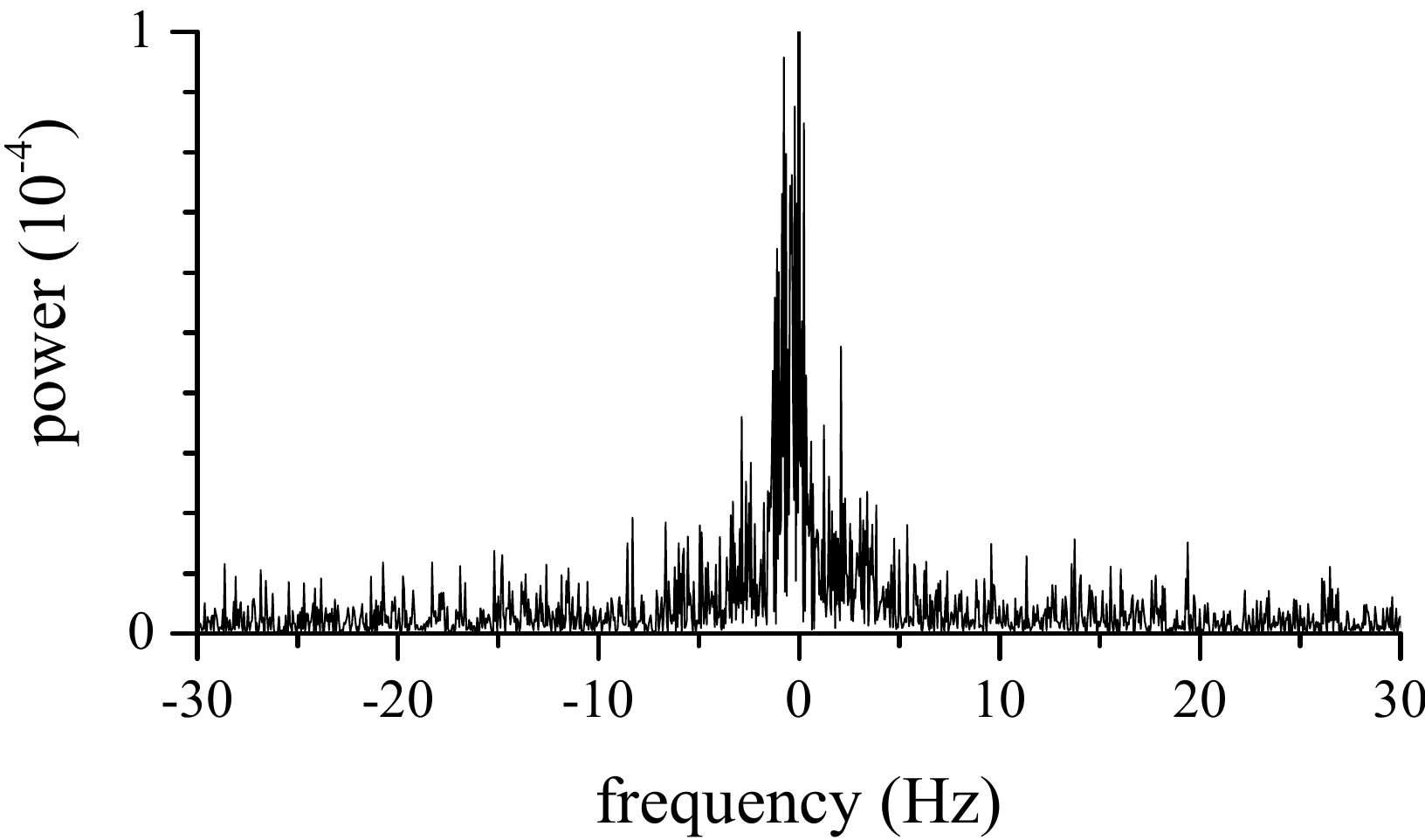}
	\caption{Power spectrum of FWM field ratio not showing the 25\,Hz
		oscillation imposed onto the sample stage for calibration, for the
		NP traces shown in the main paper Fig.7c-f.\label{powerspectrum}}
\end{figure*}

For practical purposes, we implemented a simple way to achieve {\it
	in situ} calibration of the in-plane NP position coordinates without
the need for prior knowledge and/or characterization of the particle
optical response. This is achieved by applying a small oscillation of
known amplitude (here 16\,nm) at 25\,Hz frequency to the $x$ axis of the sample
stage, which is accurately measured by the position sensor in the
stage. When the FWM field ratio from the NP encodes the NP in-plane
position as discussed in the main manuscript, this oscillation is
detected in the measured $\EF^- / \EF^+$. To find the amplitude
$\widetilde{A}$ of this oscillation (as a complex number i.e.
including its phase) we then Fourier transform $\EF^- / \EF^+$ and
multiply it with a Gaussian filter centered at 25\,Hz (with 0.5\,Hz
FWHM). We also Fourier transform the detected $x$ position sensor
from the sample stage to deduce the corresponding complex amplitude
$\widetilde{x}_0$. We can then calibrate $\EF^- / \EF^+$ in terms of
NP in-plane positions as $X={\rm
	Re}((\widetilde{x}_{0}/\widetilde{A})\EF^- / \EF^+)$ and $Y={\rm
	Im}((\widetilde{x}_{0}/\widetilde{A})\EF^- / \EF^+)$.

Conversely, if a NP exhibits a FWM field ratio dominated by the
shape asymmetry, we do no longer observe this oscillation in $\EF^-
/ \EF^+$. This case is shown in the paper in Fig.7c-f.  The power
spectrum for this case is shown in Fig.\,\ref{powerspectrum}.

\end{document}